\documentclass{aa}  
\usepackage{graphicx}
\usepackage{comment}
\usepackage{placeins}
\usepackage{txfonts}
\usepackage{multicol}
\usepackage{soul,xcolor}
\def\ramses    {{\sc ramses}}
\def\rascas    {{\sc rascas}}
\def\sphinx    {{\sc sphinx}}
\def\sphinxcr    {{\sc sphinx-cr}}
\def\sphinxsn    {{\sc sphinx-sn}}

\def\spice    {{\sc spice}}

\def\ramsesrt    {{\sc ramses-rt}}
\def\adaptahop   {{\sc adaptahop}}

\newcommand{\HI}{\textsc{{\rm H}\kern 0.1em{\sc i}}}
\newcommand{\HII}{\textsc{{\rm H}\kern 0.1em{\sc ii}}}
\newcommand{\HeI}{\textsc{{\rm He}\kern 0.1em{\sc i}}}
\newcommand{\HeII}{\textsc{{\rm He}\kern 0.1em{\sc ii}}}
\newcommand{\HeIII}{\textsc{{\rm He}\kern 0.1em{\sc iii}}}

\newcommand{\nocr}{strong SN}
\newcommand{\crs}{strong CR}
\usepackage[colorlinks=true, citecolor=blue]{hyperref}
\usepackage{natbib}

\begin{document}

   \title{The impact of cosmic ray feedback during
   the epoch of reionisation}

   \author{Marion Farcy \inst{1}\thanks{E-mail: marion.farcy@epfl.ch}
          \and Joakim Rosdahl\inst{2}
          \and Yohan Dubois\inst{3}
          \and Jérémy Blaizot\inst{2}
          \and Sergio Martin-Alvarez\inst{4}
          \and Martin Haehnelt\inst{5}
          \and Taysun Kimm\inst{6}
          \and Romain Teyssier\inst{7,8}
          }

   \institute{
   Institute of Physics, Laboratory for Galaxy Evolution, EPFL, Observatoire de Sauverny, Chemin Pegasi 51, 1290 Versoix, Switzerland
   \and Centre de Recherche Astrophysique de Lyon, CNRS UMR 5574, Univ. Lyon, Ens de Lyon, 9 avenue Charles André, F-69230 Saint-Genis-Laval, France
   \and Institut d’Astrophysique de Paris, CNRS, UMR 7095, Sorbonne Université, 98 bis bd Arago, 75014 Paris, France
   \and Kavli Institute for Particle Astrophysics \& Cosmology (KIPAC), Stanford University, Stanford, CA 94305, USA
   \and Kavli Institute for Cosmology and Institute of Astronomy, Madingley Road, Cambridge, CB3 0HA, UK
   \and Department of Astronomy, Yonsei University, 50 Yonsei-ro, Seodaemun-gu, Seoul 03722, Republic of Korea
   \and Department of Astrophysical Sciences, Princeton University, Peyton Hall, Princeton, NJ 08544, USA
   \and Program in Applied and Computational Mathematics, Princeton University, Fine Hall Washington Road, Princeton NJ 08544-1000 USA
   }

   \date{Received X X, 2025; accepted X X, 2025}

  \abstract{Galaxies form and evolve via a multitude of complex physics. In this work, we investigate the role of cosmic ray (CR) feedback in galaxy evolution and reionisation, by examining its impact on the escape of ionising radiation from galaxies. For this purpose, we present two \sphinx{} cosmological radiation-magneto-hydrodynamics simulations, enabling, for the first time,  a study of the impact of CR feedback on thousands of resolved galaxies during the Epoch of Reionisation (EoR). The simulations differ in their feedback prescriptions: one adopts a calibrated strong supernova (SN) feedback, while the other  reduces the strength of SN feedback and includes CR feedback instead. We show that both  regulate star formation and match observations of high-redshift UV luminosity functions to a reasonable extent, while also producing a similar amount of hydrogen ionising photons. In contrast to the model with strong SN feedback, the model with CRs lead to incomplete reionisation, which is in strong disagreement with observational estimates of the reionisation history. This is due to CR feedback shaping the ISM differently, filling with gas the low-density cavities carved by SN explosions. As a result, this reduces the escape of ionising photons, at any halo mass, and primarily in the close vicinity of the stars. Our study indicates that CR feedback regulates galaxy growth during the EoR, but negatively affects reionisation. This tension paves the way for the further exploration and refinement of existing galaxy formation and feedback models. Such improvements are crucial in capturing and understanding the process of reionisation and the underlying evolution of galaxies through cosmic time.}


   \titlerunning{Cosmic rays during the epoch of reionisation} \authorrunning{M. Farcy et al.}\maketitle

\section{Introduction}

The first billion years of the Universe are marked by a major phase transition: the reionisation of the inter-galactic medium (IGM). This Epoch of Reionisation (EoR, see e.g. \citealp{Barkana&Loeb2001, Zaroubi2013, Gnedin&Madau2022} for reviews), begins with the formation of the first galaxies. The IGM, initially neutral after the recombination around redshift $z=1100$, is transformed by the ultra-violet (UV) photons emitted by stars \citep{Shapiro&Giroux1987,Madau1999,Finkelstein2015} and, to a lesser extent, quasars \citep{Grazian2018,Kulkarni2019,Mason2019b,Trebitsch2021,Asthana2024}. The hydrogen-ionising radiation they emit -- also called Lyman continuum (LyC) radiation -- progressively photoionises their environment, creating growing bubbles of ionised hydrogen. By $z=5$, the hydrogen in the IGM has entirely transitioned to an ionised state \citep{Gaikwad2023}, where it remains at the present day. 

The reionisation of the IGM is a very complex and inhomogeneous process, which is described by the production of ionising photons, recombinations, and the escape fraction, $f_{\rm esc}$, of these ionising photons from the ISM of galaxies where they are produced. Star-forming galaxies are the most promising candidates for producing the LyC photons responsible for the reionisation of the Universe (\citealp{Haardt&Madau2012,Dayal2020,Dayal2024,Yeh2023,Atek2024}, but see also \citealp{Madau2024} for a recent discussion about the role of quasars). However, it remains to be determined whether the bulk of the ionising photon budget comes from the bright, but rare galaxies (so-called reionisation by the oligarchs, e.g. \citealp{Naidu2020}) or from the multitude of faint ones (also referred to as democratic reionisation, e.g. \citealp{Finkelstein2019}). Disentangling the two scenarios is especially difficult because of the faintness of potential sources of reionisation \citep{Bian&Fan2020,Mestric2020} and also due to the opacity of the IGM at high redshift \citep{Inoue&Iwata2008,Steidel2018,Bassett2021}, making it hard to provide accurate LyC emission estimates.

The escape fraction is another very important factor in understanding (and modelling) reionisation. Observationally, $f_{\rm esc}$ is highly challenging to infer (if not simply impossible) due to the significant presence of neutral hydrogen at $z>6$. While $f_{\rm esc}$ can be estimated from observations at lower redshift \citep[e.g.][]{Flury2022}, galaxies significantly leaking ionising radiation are rare \citep{Nestor2013,Japelj2017,Naidu2018,Kerutt2024}, and the values of $f_{\rm esc}$ may not be representative of those during the EoR \citep[e.g.][]{Robertson2013}. Indeed, $f_{\rm esc}$ likely varies with redshift and galaxy mass \citep{Saldana-Lopez2023,Lin2024}, which makes it even trickier to rely on low-redshift measurements only. 

Numerical simulations are our best chance to understand which physical processes and which galaxies regulate the production and escape of LyC photons, and, ultimately, to interpret observations of the EoR. Among the surge of radiation-hydrodynamics (RHD) cosmological simulations, most of them focus on the overall process of reionisation \citep[e.g.][]{Gnedin2014,Pawlik2017,Finlator2018,Ocvirk2020,Kannan2022}, at the cost of capturing the small scales at which radiation is emitted and propagated. This limits their predictive power when it comes to understand which physical processes actually rule the reionisation. Achieving this demands simulations that model the multi-phase inter-stellar medium (ISM) and do not consider $f_{\rm esc}$ as an input parameter. For this purpose, one common approach is to rely on cosmological zoom simulations \citep{Wise2014,Ma2015,Kimm&Cen2014,Paardekooper2015,Kimm2017,Trebitsch2017,Trebitsch2021,Lovell2021}. This type of simulation provides insights into the physics governing the escape of ionising radiation, at the sacrifice of modelling the IGM, unlike non-zoomed cosmological simulations that reconcile IGM and ISM scales studies of the EoR \citep{OShea2015, Rosdahl2018,Bhagwat2024}. In this paper, we focus on the framework of the \sphinx{} simulations \citep{Rosdahl2018,Rosdahl2022,Katz2023}.

Despite their diversity in size, resolution, and galaxy formation models, simulations of the reionisation all show that $f_{\rm esc}$ is very much controlled by the physical processes that regulate star formation and the ejection of gas, clearing the way for radiation to propagate from galaxies to the IGM. This picture is also supported by observations of LyC-leaking galaxies, which emphasise the crucial role of stellar feedback from young stars in facilitating the escape of LyC photons \citep[e.g.][]{Mainali2022,Carr2025,Flury2024}. Feedback from supernova (SN) explosions is probably the most promising channel for carving low-density pathways through which LyC photons can efficiently escape \citep{Kimm&Cen2014,Rosdahl2018,Cen2020,Gazagnes2020,Saldana-Lopez2022}. Other processes operating on more continuous timescales, such as turbulence and radiation feedback, can also play a similar role by impacting the geometry and structure of the ISM \citep{Kim2018,Kakiichi&Gronke2021,Carr2025,Jaskot2024b}.

Simulations are a valuable tool for investigating the contribution of these different physical processes in shaping galaxies and the escape of LyC photons. However, they remain limited by numerical uncertainties in the modelling of feedback processes \citep[see e.g.][for how different models of SN feedback impact the reionisation of the IGM]{Bhagwat2024}. To capture the impact of feedback below the resolution scale, cosmological simulations have to make use of subgrid models. These models are often empirically calibrated, in order to overcome the lack of resolution and physics, and to reproduce realistic galaxy properties \citep[e.g.][]{Teyssier2013,DallaVecchia&Schaye2012,Rosdahl2018,Semenov2018}. Now that it is within reach to model feedback from first principles, we can draw our attention to other complementary mechanisms that are missing in our models of galaxy evolution. 

One such promising important source of feedback comes from cosmic rays (CRs), which are charged particles that are thought to be accelerated at shocks \citep{Axford1977,Bell1978,Blandford&Ostriker1978}. CR feedback has already been shown to regulate star formation and to contribute to remove dense gas out of the ISM, impacting galaxy properties at ISM and circum-galactic medium (CGM) scales \citep[e.g.][]{Jubelgas2008,Booth2013,Salem&Bryan2014,Butsky&Quinn2018,Girichidis2018,Hopkins2020,Dashyan&Dubois2020,Martin-Alvarez2023,Curro2024}. In addition, \citet{Farcy2022} showed that CR feedback in idealised disc galaxies reduces the escape of hydrogen ionising radiation, but the consequences on the reionisation process in a more realistic cosmological context, are yet to be determined.

In this study, our goal is to address a number of questions unexplored until now, namely:\ whether  CRs an important source of feedback in the early Universe; how they impact the growth of galaxies during the EoR; whether they affect the propagation of the LyC radiation and, if so, what role they play in the reionisation of the Universe. To answer these questions, we perform and study two $\left(10 \,\rm Mpc\right)^3$ \sphinx{} cosmological radiation-magneto-hydrodynamical (RMHD) simulations, run down to $z=5$. Both simulations include SN feedback, and one models CR injection from SNe and transport via anisotropic diffusion. This paper therefore introduces the first non-zoomed CR-RMHD cosmological simulation to date, which allows us to study the effects of CR feedback on reionisation and in thousands of high-redshift galaxies.

We structure this paper as follows. In Sect.~\ref{section:simu} we present the details of the two \sphinx{} simulations used in this study, along with their two calibrated SN and CR feedback models. To illustrate the effect of CR feedback, Sect.~\ref{section:results} starts by showing qualitative visualisations of the simulations at different scales. We then demonstrate in Sect.~\ref{subsection:sfr} that our simulations with and without CR feedback both lead to a sufficient regulation of star formation, producing galaxies with stellar masses and UV luminosities calibrated to match the realistic UV luminosity functions from the original \sphinx{} simulations. Section~\ref{subsection:eor} focuses on the global impact of SN and CR feedback on the reionisation history and Sect.~\ref{subsection:fesc} investigates in more details in which halos and at which scales CR feedback suppresses the escape of LyC radiation. We discuss the implications of CR feedback on reionisation in Sect.~\ref{section:disc} and summarise our results in Sect.~\ref{section:ccl}.

\section{Simulations and methods}
\label{section:simu}

To study the effect of CR feedback on reionisation and early galaxy evolution, we performed two \sphinx{} RMHD cosmological simulations. These simulations (and the code used to run them) are very similar to those presented in \citet{Rosdahl2018}. One of the main differences between the fiducial \sphinx{} simulation and those presented in this paper is the inclusion of a magnetic field together with the use of a magnetohydrodynamic solver, instead of a purely hydrodynamic one. As described in what follows and as done in \citet{Rosdahl2022} for the largest and most recent \sphinx{} simulation to date, we also employed an updated spectral energy distribution (SED) model, two radiation groups (instead of three), and single precision RHD. For one of our two simulations, we included CR injection, transport, and feedback, parametrised to similarly regulate star formation and UV luminosities as in the counterpart run that does not include CRs. We summarise below the main characteristics of the simulations. We refer to \citet{Rosdahl2018} and \citet{Rosdahl2022} for a more complete description of the fiducial \sphinx{} set of simulations.

\subsection{Simulation code and initial conditions}
\label{subsection:ics}

\subsubsection{Code and solvers}

The \sphinx{} simulations were performed using the \ramses{} adaptive mesh refinement (AMR) code \citep{Teyssier2002}. To track the non-equilibrium ionisation state of the gas, we used its radiation-hydrodynamics extension, \ramsesrt{}, based on the first-order moment radiative transfer method implemented by \citet{Rosdahl2013, Rosdahl&Teyssier2015}. We solved the ideal MHD equations with the Harten-Lax-van Leer discontinuities (HLLD) Riemann solver \citep{Miyoshi&Kusano2005} and the minmod total variation diminishing slope limiter \citep{vanLeer1979}, as implemented by \citet{Fromang2006}. The induction equation for the magnetic field is solved with a constrained transport method, following the MUSCL second order Godunov scheme \citep{Teyssier2006}. Here, gas is considered as a purely monoatomic ideal gas and modelled with an adiabatic index $\gamma=5/3$. To account for CR anisotropic diffusion, we used the solver developed by \citet{Dubois&Commercon2016}, together with the minmod slope limiter on the transverse component of the flux,  to preserve the monotonicity of the solution, as described in \citet{Dashyan&Dubois2020}.

\subsubsection{Initial conditions}

In this work, we used the same initial conditions (ICs) as the $\left(10\,\rm cMpc\right)^3$ \sphinx{} simulation of the suite \citep{Rosdahl2018}. The ICs of the \sphinx{} simulations are generated with the {\sc{music}} code \citep{Hahn&Abel2011}. They start at $z=150$ and are chosen as  the average among a set of 60 ICs in order to minimise the effect of cosmic variance. The \sphinx{} simulations follow a $\Lambda$CDM Universe, whose cosmological parameters are adopted from the results of \citet{Planck2014A1}. In particular, the total matter density is $\Omega_{\rm m}=0.3175$, the cosmological constant density is $\Omega_{\rm \Lambda}=0.6825$, the baryon density is $\Omega_{\rm b}=0.049,$ and the Hubble constant is $H_{0}=67.11\,\rm km\,s^{-1}\,Mpc^{-1}$. The simulations have a primordial hydrogen mass fraction $X=0.76$, a helium mass fraction $Y=0.24$, and assume an initial homogeneous metal gas fraction $Z_{\rm ini}=3.2\times10^{-4}\rm\, Z_\odot$ (assuming that the Solar metal mass fraction is $\rm Z_\odot=0.02$) to form the first stars at $z\approx15$, compensating for the lack of primordial molecular hydrogen cooling in the simulations\footnote{The exact value has been calibrated to reproduce the timing and efficiency of star formation from zoom simulations that include primordial molecular hydrogen formation and cooling, and that adopt a similar physical set-up as the \sphinx{} simulations otherwise \citep{Kimm2017,Rosdahl2018}.}.

\subsubsection{Initialisation of the magnetic field}

We initialised the magnetic field as a random field, by creating a random Gaussian vector potential field over a uniform grid, with a coherence length of 39 ckpc. The magnetic field components at the interfaces of each cell were then reconstructed from the curl of the potential, such that the magnetic field could remain divergence-free. The magnetic field of each cell was normalised such that the initial magnetic field has a magnitude of $10^{-12}$ comoving Gauss \citep{Shaw&Lewis2012,Planck2016}, which was not expected to have a significant effect in reionisation \citep{Katz2021}. Given this initialisation, the magnetic field reaches a strength of $0.01-1 \rm\, \mu G$ in the ISM of galaxies soon after their formation.

\subsection{General set-up}
\label{subsection:setup}

\subsubsection{Resolution and refinement strategy}

The \sphinx{} simulations studied here have volumes of $\left(10\,\rm cMpc\right)^3$. They are composed of $512^3$ collisionless dark matter (DM) particles, each with a mass of $2.5\times10^{5}\rm\,M_\odot$. They have a fixed co-moving resolution, with minimum and maximum cell widths that are 76.3 cpc and 19.5 ckpc, respectively. As the refinement levels are fixed throughout the simulations, the resolution becomes lower with decreasing redshift, with the minimum and maximum cell widths reaching 12.7 pc and 3.3 kpc at $z=5$, respectively. We followed an adaptive refinement strategy, where a cell is refined if the sum of its dark matter mass and a fraction $\Omega_{\rm m}/\Omega_{\rm b}$ of its baryonic mass is higher than the mass of eight DM particles or if its width is larger than a quarter of the local Jeans length (provided that gas density is higher than $30\,\rm H\,cm^{-3}$ to avoid prohibitive refinement in the IGM).

\subsubsection{Radiation and non-equilibrium chemistry}

Radiation injection, propagation, and interaction with gas via photoionisation, heating, and momentum transfer are described in \citet{Rosdahl2013} and \citet{Rosdahl&Teyssier2015}. To speed up the calculation of the radiative transfer equations, we used the variable speed of light approximation as described in \citet{Katz2017}, such that the speed of light goes from 1.25 per cent to 20 per cent of the real value in the highest to lowest resolution cells, respectively. We split radiation into two photon groups, which correspond to hydrogen and helium ionising photons. We tracked the non-equilibrium abundances of neutral hydrogen and helium and of \HII{}, \HeII{} and \HeIII{}. 

Stellar particles emit radiation as a function of their ages and metallicities, as derived from version 2.2.1 of the Binary Population And Spectral Synthesis model (BPASS, \citealp{Stanway2016,Stanway&Eldridge2018}). The SED model assumes an initial mass (IMF) function close to \citet{Kroupa2001} with slopes of -1.3 from 0.1 to 0.5 $\rm M_{\odot}$ and -2.35 from 0.5 to 100 $\rm M_{\odot}$. 

The non-equilibrium thermochemistry of hydrogen and helium is described in \citet{Rosdahl2013}, accounting for collisional ionisation, photoionisation, collisional excitation, (dielectric) recombination, bremsstrahlung emission, and inverse Compton scattering off of cosmic microwave background photons. In addition, we included atomic metal cooling for gas at temperature of $T>10^4\,\rm K,$ via cooling rate tables pre-calculated with \textsc{cloudy} \citep{Ferland1998}, and fine structure metal cooling for gas at $15\,{\rm K}\leq T\leq 10^4\,\rm K$, adopting the rates from \citet{Rosen&Bregman1995}.

\subsubsection{Star formation}

For star formation, we adopted the same gravo-turbulent model as in \citeauthor{Rosdahl2018} (\citeyear{Rosdahl2018}, see also \citealp{Kimm2017} or \citealp{Trebitsch2017} for details). We note that this star formation model does not include any contribution from the CR pressure to facilitate the comparison between the two simulations studied in this paper. Gas is stochastically converted into stellar particles \citep{Rasera&Teyssier2006} following a Schmidt law, with a local efficiency based on the gravo-turbulent properties of the gas \citep{Hennebelle&Chabrier2011, Padoan&Nordlund2011,Federrath&Klessen2012}. In particular, this star formation efficiency varies with the local virial parameter of $\alpha_{\rm vir}=2E_{\rm k}/E_{\rm g}$ (where $E_{\rm k}$ and $E_{\rm g}$ are  the turbulent and gravitational energies of the gas, respectively) and increases with the turbulent Mach number in gravitationally well-bound regions ($\alpha_{\rm vir}\lesssim1$, for details, see Eqs. 2 and 3 from \citealp{Kimm2017} and Fig. 1 from \citealp{Federrath&Klessen2012}). In order to be turned into a stellar particle with an initial mass of $\rm 1000\,M_\odot$, gas has to be locally convergent and reside in cells at the highest level of refinement, with a size larger than the turbulent Jeans length. We could also prevent a star particle from forming if it ended up removing more than 90 per cent of the gas mass of its candidate host cell.

\subsubsection{Supernova feedback}

As in \citet{Rosdahl2018}, we modelled type II SN explosions using the mechanical feedback model described in \citet{Kimm&Cen2014} and \citet{Kimm2015}. This model ensures the accurate transfer of radial momentum to the surrounding medium by distinguishing between adiabatic and momentum-conserving phases. Stellar particles undergo multiple SN explosions sampled between 3 and 50 Myr after their birth, each explosion releasing an energy $E=10^{51}\rm erg$. The fraction of mass recycled into SN ejecta is 20\% (close to a \citealp{Kroupa2001} IMF) and 7.5\% of this mass is recycled back into the ISM as elements heavier than hydrogen and helium. In the fiducial \sphinx{} RHD simulations, the number of SN explosions per solar mass is set to 4 SNe per $100\,\rm M_\odot$ formed, which roughly corresponds to four times the number derived from the \citet{Kroupa2001} IMF. This artificial calibration of the rate of SN explosions has been adopted in order to reproduce realistic high-redshift galaxies, in terms of UV luminosity function \citep{Rosdahl2018,Rosdahl2022,Katz2023}. This boost in SN feedback, which has the effect of suppressing star formation, can be interpreted as accounting for a number of modelling uncertainties, such as the lack of resolution, which can lead to numerical overcooling of the gas and hence enhanced star formation, and physical processes unaccounted for in the simulations, such as CR feedback, which we want to investigate in this paper. Boosting SN feedback can also be interpreted as the consequence of a top-heavy IMF, which may or may not be prevalent at high redshift \citep[e.g.][]{Cameron2024}, or a higher energy injection from SN explosions, such as that characterising hypernova events \citep[e.g.][]{Kobayashi2006}. In this work, we adopt this boosted SN feedback in one of our two RMHD \sphinx{} simulations (labelled the \sphinxsn{} simulation). We refer to this feedback as the \nocr{} model. In the other simulation, we reduced this artificial calibration by a factor of two (meaning 2 SNe per $100\,\rm M_\odot$), substituting the weaker SN feedback with CR feedback.

\subsubsection{Cosmic ray feedback}

In our \sphinx{} simulation with CR feedback (referred to as the \sphinxcr{} simulation), we modelled the CR advection with the bulk motion of the gas and anisotropic diffusion along magnetic field lines, as described in \citet{Dubois&Commercon2016} and \citet{Dubois2019}. In \ramses{}, CRs are treated as a non-thermal pressure from a relativistic fluid, with an adiabatic index $\rm \gamma_{CR}=4/3$. Considering CRs as collisionless particles of a few $\rm GeV$, we used a diffusion coefficient $\kappa=10^{28}\,\rm cm^2\,s^{-1}$ \citep{Strong2007, Trotta2011} and accounted for radiative (hadronic and Coulomb) CR energy losses following \citet{Guo&Oh2008}. We neglected the effect of the CR streaming instability in these simulations and we defer a study of streaming to future work. We should also note that in our study, the thermal momentum in the SN snowplough phase does not include any CR energy density, unlike the approach in \citet{Diesing&Caprioli2018} and \citet{Curro2022}.

\begin{table}
    \caption{Differences between the two \sphinx{} simulations analysed in this paper.}
    \centering
    \begin{tabular}{cccc} \hline
        Simulation name & Feedback name & \# / 100 $\rm M_\odot$ & $f_{\rm ecr}$ \\ \hline
        \sphinxsn{} & \nocr{} & 4 & 0 \\ 
        \sphinxcr{} & \crs{} & 2 & 0.2 \\ 
        \hline
    \end{tabular}
    \tablefoot{From left to right, columns are: name of the simulation, name of the feedback model, \# / 100 $\rm M_\odot$: number of SN explosions per $\rm 100\, M_\odot$ stellar mass formed, $f_{\rm ecr}$: fraction of SN energy injected into CR energy.}
    \label{tab:simulations}
\end{table}

Shock waves generated by SN explosions can efficiently accelerate CRs \citep{Axford1977,Krymskii1977,Bell1978,Blandford&Ostriker1978}. Usually, simulations of CR feedback that do not resolve these SN shocks inject 10 per cent of the SN energy into CRs (e.g. \citealp{Pfrommer2017,Dashyan&Dubois2020,Hopkins2020}, but see also \citealp{Jubelgas2008,Butsky&Quinn2018,Semenov2021}), based on observations of local SN remnants \citep{Hillas2005, Strong2010,Morlino&Caprioli2012,Dermer&Powale2013}. However, studies diverge, finding values that can reach up to 40 per cent \citep{Kang&Jones2005, Ellison2010,Helder2013,Bhadra2022}, which may translate the fact that SN explosions occur in places where CRs have already been injected, so that this pre-existing CR population is further accelerated at a higher rate than the canonical 10 per cent \citep{Caprioli&Spitkovsky2014,Caprioli2018, Vieu2022}. 

In this work, we want to calibrate our two \sphinx{} simulations (with and without CRs) such that they match the high-redshift UV luminosity functions in a similar way as the original \sphinx{} simulations. For this purpose, we chose to inject 20 per cent of SN energy into CRs, which remains within the range of acceptable values, and set the number of SN explosions per $100\,\rm M_\odot$ of stars formed to 2. We refer to this feedback model, used in this \sphinxcr{} simulation, as the \crs{} feedback. In Sect.~\ref{section:disc} we offer more details on our choice of SN rate and CR energy injection values, which have been chosen among several variations based on a series of smaller \sphinx{} simulations (to be presented in a forthcoming follow-up paper). The differences between the two simulations are summarised in Table~\ref{tab:simulations}. At each SN explosion, the non-thermal CR energy was injected into the host cell, such that the total energy from both the SN and CRs is $E=10^{51}\rm erg$. 

\subsection{Halo identification}
\label{subsection:halo}

\begin{figure}
        \includegraphics[width=\columnwidth]{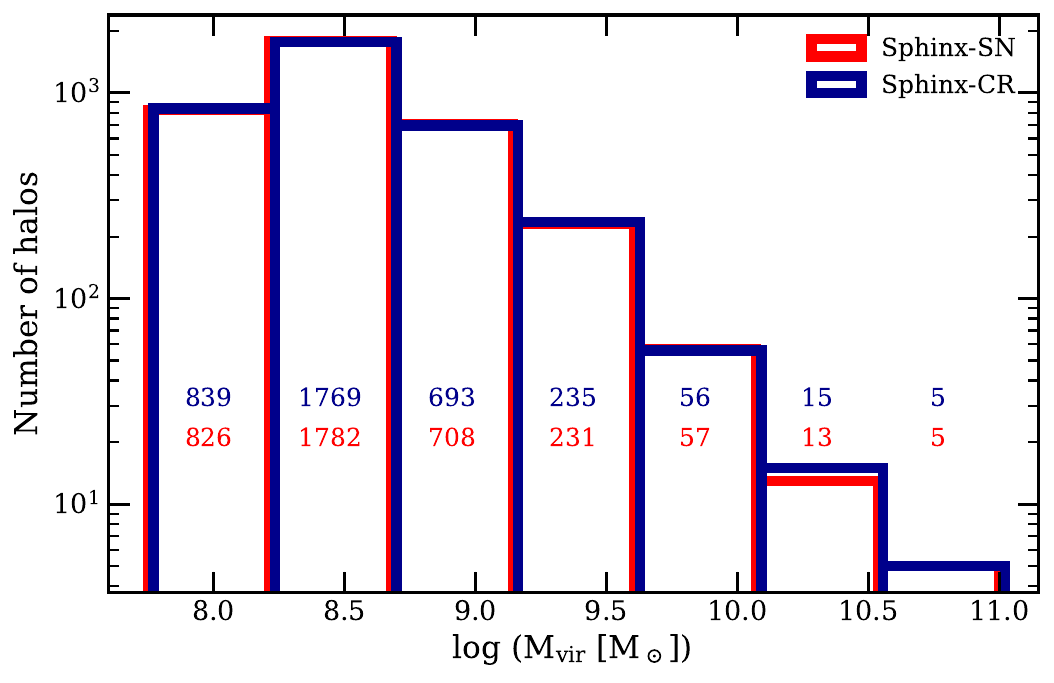}
        \centering
    \caption{Histogram of the number of $z=5$ dark matter halos per virial mass for \sphinxsn{} (in red) and \sphinxcr{} (in dark blue). The number of halos in each bin is written with the same colour code, and roughly corresponds to a total of 3600 resolved halos (i.e. with a DM mass higher than $7.5\times10^7\rm\,M_\odot$) for both simulations.}
    \label{fig:hist}
\end{figure}

To identify DM halos, we used the \adaptahop{} algorithm in the most massive submaxima mode \citep{Aubert2004,Tweed2009}, as in \citet{Rosdahl2018}. A halo is defined as a region in which the virial theorem is satisfied and that contains at least 20 DM particles. In this work, we have ignored the sub-halos and only considered the main resolved halos, defined as those enclosing 300 DM particles so that they have a minimum mass of $M_{\rm vir}=7.5\times10^7\rm\,M_\odot$, which is above the atomic cooling limit \citep{Wise2014}. Figure~\ref{fig:hist} shows the mass distribution of halos in the two simulations at $z=5$, which illustrates the number of star forming halos per bin of virial mass. The two simulations have a similar number of halos, which is 3621 for \sphinxsn{} and 3611 for \sphinxcr{} at $z=5$. These small differences are expected to be mostly due to subtleties of the halo finder. The virial mass of the most massive halo is $6.9\times10^{10}\,\rm M_\odot$ at $z=5$.

\subsection{Lyman continuum escape fractions}
\label{subsection:rascas}

The radiative transfer method implemented in \ramsesrt{} allows us to determine the ionisation state of the gas, and models the effects of radiation on gas. However, individual photons and their direction of propagation are not tracked; thus, it is not directly possible to determine where and when the radiation emitted by a stellar particle is absorbed. Furthermore, an indirect determination of $f_{\rm esc}$ using the radiation field evolved in the simulation is complicated due to the reduced and variable speed of light. Therefore, to compute the escape fractions of LyC photons for each halo, we used the public radiative transfer code \rascas{} \citep{rascas2020} and followed a procedure that is similar to the one described by \citet{Rosdahl2022}. For each snapshot, the halos were first identified with the halo finder algorithm described previously. For each halo, photon packets were then cast isotropically from stellar particles with a probability proportional to their LyC luminosity and were propagated following a Monte-Carlo procedure described by \citet{rascas2020}. We considered  the LyC photons to be propagating until they are absorbed by neutral hydrogen or helium\footnote{For consistency with the LyC absorption done on-the-fly in the simulations, we neglect dust absorption when computing the escape fractions of LyC photons with \rascas{}. The same choice has been made in the introductory \sphinx{} papers from \citet{Rosdahl2018} and \citet{Rosdahl2022}, and \citet{Kimm2019} showed that the effect of dust absorption on LyC escape fractions is subdominant compared to absorption by hydrogen.}, which occurs with a probability that depends on the optical depth; that is to say, on the cross-section of interaction between neutral hydrogen and LyC photons, as well as on the column density of neutral hydrogen along the line of sight. To determine the escape fraction associated to one halo, the number of LyC photons that reach the virial radius boundary of the halo without being absorbed is compared to the total number of LyC photons emitted by all the stellar particles of the halo, derived from the SED used as a function of the age and metallicity of each stellar particle. The global escape fraction is eventually measured as the intrinsic LyC luminosity-weighted escape fraction for all the rays in the volume sampled by \rascas{}.

\subsection{UV magnitude and luminosity}
\label{subsection:muv}

To compare the luminosities of the galaxies formed in the \sphinx{} simulations to observations, we also use \rascas{} to compute the magnitude of each halo at $1500\,\rm \mathring{A}$, otherwise known as the UV luminosity. In this case, the same procedure described to compute the escape fraction of LyC is used, additionally taking into account the effect of dust. The UV photons can be absorbed and scattered by dust grains with a probability scaling with the dust albedo $A=0.38$, following \citet{Li&Draine2001}. Dust grains are not directly modelled with \ramses{} nor \rascas{}. Instead, \rascas{} models the dust absorption in each cell, defining a dust absorption coefficient depending linearly on the cell gas metallicity, its neutral and ionised hydrogen densities, and on the dust cross-section per atom of hydrogen \citep[as also described by][]{Garel2021}. The latter is normalised to the extinction curve of the Small Magellanic Cloud (which is appropriate for low-mass high-redshift galaxies that have young stellar populations) following \citet{Laursen2009} and \citet{Smith2018}.

Following this prescription gives us the intrinsic $1500\,\rm \mathring{A}$ luminosity and the escape fraction of the corresponding photons that have not been absorbed by dust. The product of these two quantities gives the dust-attenuated UV luminosity $L_{1500}$, which is eventually converted into the magnitude $M_{1500}$ following the definition in \citet{Oke&Gunn1983}:

\begin{equation}
    \label{eq:MAB}
    M_{1500} = 51.595 -2.5\log\left(\frac{L_{1500}}{\rm erg\,s^{-1}\,Hz}\right)\,.
\end{equation}

\section{Results}
\label{section:results}

\begin{figure*}
        \includegraphics[width=0.95\textwidth]{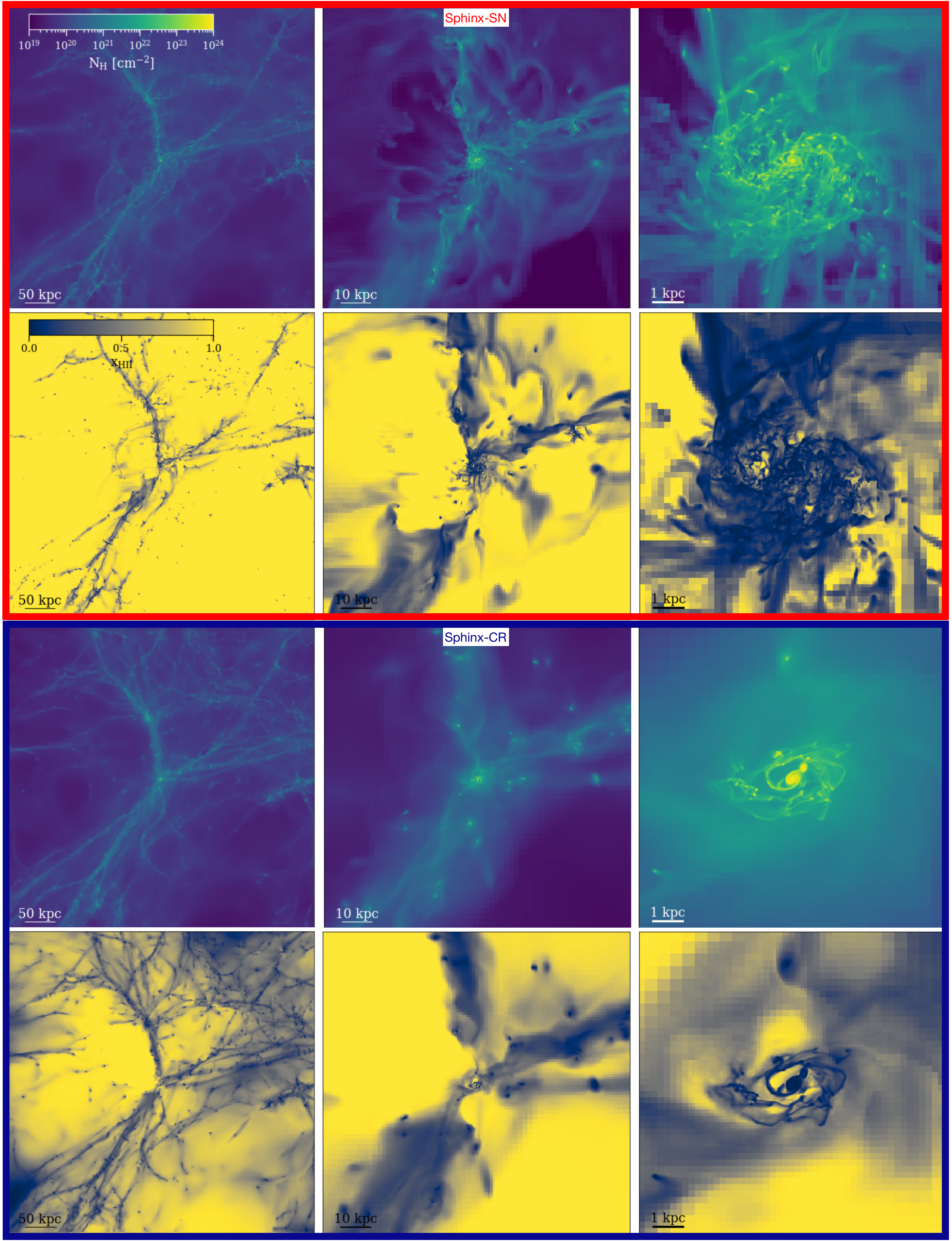}
        \centering
    \caption{Hydrogen gas column density (first rows) and mass-weighted fraction of ionised hydrogen (second rows) maps centred on the most massive halo at $z=5$ from \sphinxsn{} (top) and \sphinxcr{} (bottom). From left to right, maps are 500, 100 and $10\,\rm kpc$ wide, and a 50, 10 and $1\,\rm kpc$ width scale bar is respectively plotted in the lower left corner of each panel.}
    \label{fig:maps}
\end{figure*}

To qualitatively visualise how the \nocr{} and \crs{} feedback impact galaxies, Fig.~\ref{fig:maps} shows hydrogen column density and mass-weighted hydrogen ionisation fraction maps, centred on the most massive halo of our two \sphinx{} simulations at $z=5$. At this redshift, the halo has a virial mass of $\rm 6.9\times10^{10}\,M_\odot$, a virial radius of $\rm 20.5\, kpc$ and a stellar mass of $\rm 2.7\times10^9\,M_\odot$ and $\rm 9.3\times10^9\,M_\odot$ in \sphinxsn{} and \sphinxcr{}, respectively. This is the same halo in both simulations, but the galaxy varies due to the different feedback models. To better illustrate this at IGM, CGM and ISM scales, the different panels in Fig.~\ref{fig:maps} have widths of 500, 100, and $10\,\rm kpc$, zooming on the central galaxy from left to right.

The leftmost panels show the filamentary structure of the cosmic web, whose gas feeds the central galaxy. This large-scale view of the simulations gives an insight on the ionisation state of the IGM in the two simulations. While a significant fraction of gas remains neutral in \sphinxcr{}, the IGM in \sphinxsn{} is predominantly ionised away from the cosmic filaments. In the latter, we can distinguish large cavities (at very low densities and completely ionised), and shells of dense gas expanding far from the central galaxies, that are remnant of gas shocked by the SN explosions. These features do not appear on the large scale gas density map of \sphinxcr{}, in which the overall gas distribution is somewhat smoother. This can be better visualised from the middle panels of Fig.~\ref{fig:maps}. In \sphinxsn{}, we clearly distinguish low-density cavities carved by SN explosions, that facilitate the escape of ionising radiation. On the other hand, the \crs{} feedback makes the galactic gas distribution more homogeneous, and leads to a denser CGM than when CRs are not included, as already found in previous works from idealised and cosmological zoom simulations \citep[e.g.][]{Girichidis2018,Dashyan&Dubois2020,Buck2020}. At smaller scales (rightmost panels), the \crs{} feedback makes the ISM less porous than the \nocr{} feedback, with fewer dense clumps.

To summarise, the \nocr{} feedback is more disruptive than the \crs{} model, creating bubbles of low-density gas around galaxies that CR pressure fills with gas otherwise. The \nocr{} feedback leads to more filamentary structures and helps to clear out several sight lines that are absent with the \crs{} feedback. The two feedback models also produce different ISM configurations, with CRs leading to a less turbulent and fragmented ISM. This can be anticipated to impact how LyC radiation can escape, and to explain why the IGM in \sphinxcr{} is only partially ionised at $z=5$. We checked similar maps of $\sim 5\times10^9\,\rm M_\odot$ halos, and note that these conclusions also hold for less massive galaxies than those shown in Fig.~\ref{fig:maps}. In what follows, we demonstrate how the two feedback models quantitatively impact star formation in galaxies and the ionisation state of the IGM.

\subsection{Regulation of star formation and UV luminosity}
\label{subsection:sfr}

\begin{figure}
        \includegraphics[width=0.95\columnwidth]{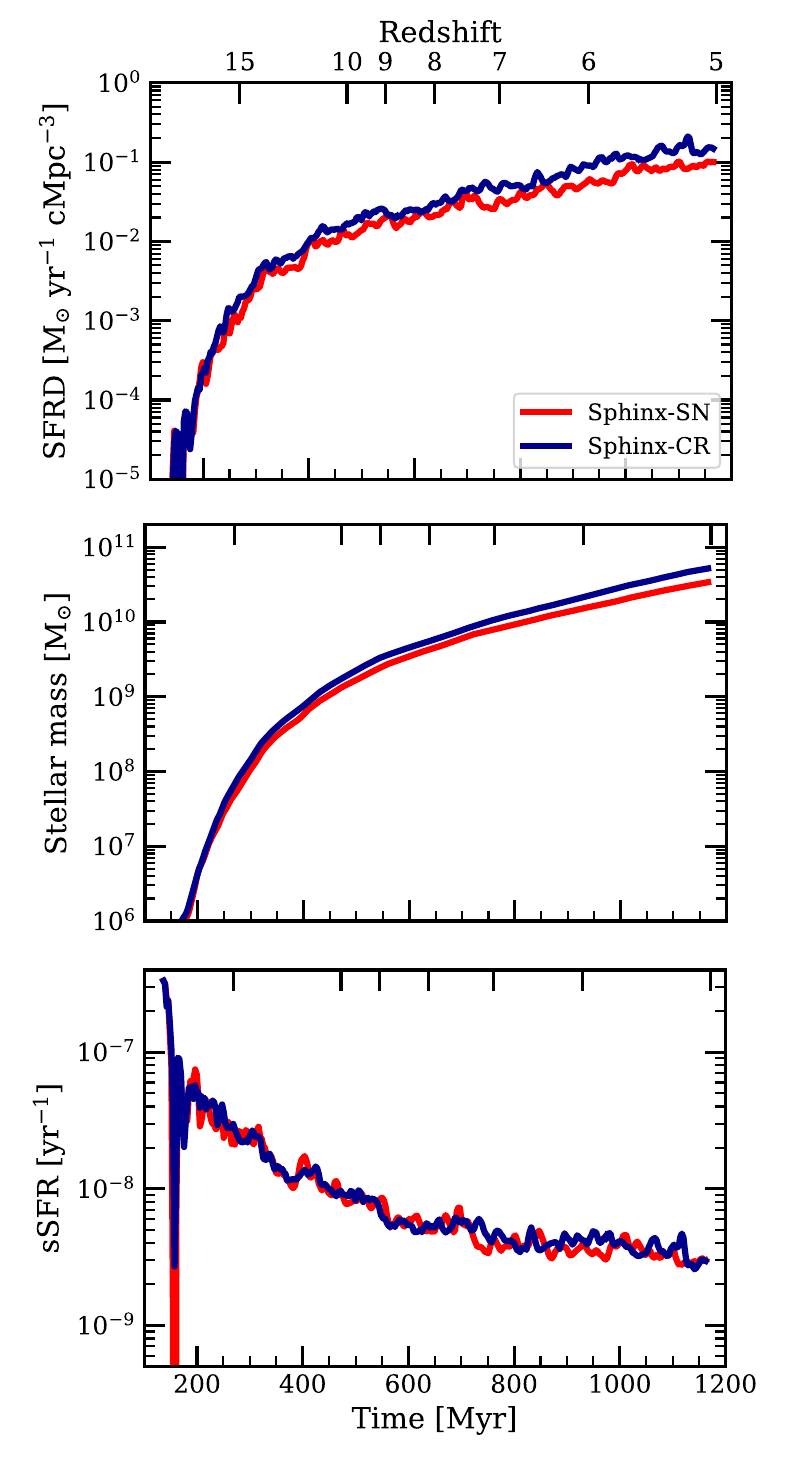}
        \centering
    \caption{From top to bottom: Time evolution of the SFRD, total stellar mass, and sSFR in \sphinxsn{} (red) and \sphinxcr{} (dark blue) runs. The SFR is calculated from all stellar particles formed in the 10 cMpc boxes, averaged over 10 Myr.}
    \label{fig:sfr}
\end{figure}

To illustrate how star formation is regulated globally in our two simulations, Fig.~\ref{fig:sfr} shows the time evolution of the star formation rate density for the simulated volume (SFRD), total stellar mass formed, and specific star formation rate ($\rm sSFR$, defined as the ratio of the SFR to the stellar mass). The SFR shown in this section is averaged over the last 10 Myr. As done throughout the paper, results from \sphinxsn{} are shown with red lines, while those from \sphinxcr{} are shown in dark blue.
At any time, the SFRD is slightly higher in \sphinxcr{} than in \sphinxsn{} but overall, the star formation histories in the two simulations are similar. In particular, they have a very similar sSFR through cosmic time. By $z=5$, the simulations have total stellar masses that differ by less than a factor of 1.5, respectively reaching $3.5\times10^{10}\,\rm M_\odot$ and $5.6\times10^{10}\,\rm M_\odot$ in \sphinxsn{} and \sphinxcr{}.

\begin{figure}
    \includegraphics[width=\columnwidth]{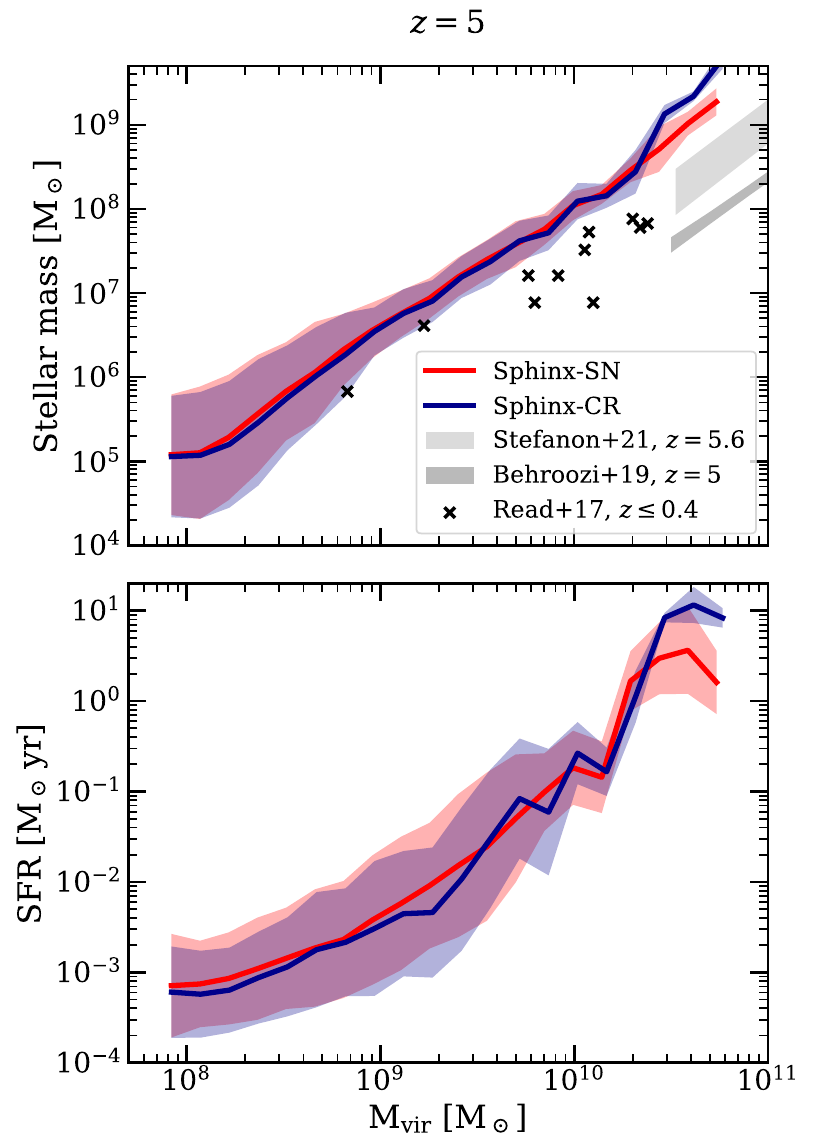}
    \centering
    \caption{SMHM relation (upper panel) and 10-Myr averaged SFR (lower panel) versus halo mass in \sphinxsn{} (red) and \sphinxcr{} (dark blue) at $z=5$. The curves respectively show the averaged stellar mass and SFR per bin of halo virial mass, and the coloured shaded regions represent the standard deviation. We also show observational constraints from \citet{Read2017} at $0.2\leq z\leq0.4$ with black crosses, from \citet{Behroozi2019} at $z=5$ with a dark grey shaded area and from \citet{Stefanon2021} at $z=5.6$ with a light grey shaded region. At any halo mass, the stellar mass is roughly the same in the two simulations, but tends to be higher than the observational constraints. At the massive end, galaxies are more massive in \sphinxcr{} and have higher SFRs.}
    \label{fig:smhm}
\end{figure}

To determine where in particular the stellar masses differ in the two simulations, the upper panel of Fig.~\ref{fig:smhm} shows the stellar-mass-to-halo-mass (SMHM) relation at $z=5$ (which corresponds to the averaged stellar mass enclosed in halos from a given virial mass bin). We additionally show the $1\sigma$ standard deviation in shaded area, which is larger at lower halo mass. At any halo mass, the stellar masses in the two simulations are very similar, which is particularly true for halos with $M_{\rm vir}\lesssim 10^{10}\rm\,M_\odot$. Conversely, the \crs{} feedback becomes less efficient at regulating star formation in the most massive halos with $M_{\rm vir}\gtrsim 3\times10^{10}\rm\,M_\odot$ (as previously found by \citealp{Jubelgas2008,Booth2013,Pfrommer2017,Jacob2018}, but see also \citealp{Chan2019,Hopkins2020}). As shown by \citet{Farcy2022}, using the same star formation and feedback models as in the \sphinx{} simulations, CR feedback suppresses star formation by reducing the number and the mass of star-forming clumps. Indeed, the CR pressure supports gas against gravitational collapse, dispersing gas locally in the ISM, and this effect is stronger in low-mass galaxies that have a shallow gravitational potential. CRs also help in driving large-scale winds, which affects the gas content of galaxies and their star formation \citep[e.g.][]{Curro2024}. Because of the small volume of our \sphinx{} simulations, there are only a few massive halos (with $M_{\rm vir}\gtrsim 3\times10^{10}\rm\,M_\odot$, see Fig.~\ref{fig:hist}) and they represent only 0.5 per cent of the halo population in number. However, massive halos contribute the most to the total stellar mass, and have the dominating contribution in the difference seen in SFRD and total stellar mass shown in Fig.~\ref{fig:sfr}.

The efficiency of CR feedback in regulating star formation can also be seen in the lower panel of Fig.~\ref{fig:smhm}, which shows the 10-Myr averaged SFR per halo mass bin at $z=5$. With the exception of the most massive galaxies, our calibrated CR feedback is as efficient at regulating star formation as the \nocr{} feedback, although the total energy released by the SN explosions is two times higher with the latter. While the \nocr{} model injects a thermal energy of $\rm 4\times10^{51}\,erg$ each $\rm 100\, M_\odot$ of stars formed, the \crs{} model injects a thermal energy of $\rm 1.6\times10^{51}\,erg$ and a CR energy of $\rm 4\times10^{50}\, erg$. Therefore, Fig.~\ref{fig:smhm} shows that CR feedback can have a similar effect on star formation as boosting the SN feedback, and that CRs may have a non negligible role in regulating galaxy growth during the EoR. 

To better assess the strength of our two feedback models, the upper panel of Fig.~\ref{fig:smhm} also shows observational estimates of the SMHM relation for local dwarf galaxies from \citet{Read2017} with black triangles, and observational constraints at $z=5$ and $z=5.6$ from \citet{Behroozi2019} and \citet{Stefanon2021} with dark and light grey shaded regions, respectively. The SMHM relations for our two simulations are slightly above the observational constraints at $z\simeq 5$ from \citet{Behroozi2019} and \citet{Stefanon2021}\footnote{We chose not to include the baryons in the calculation of the halo virial masses, as done by \citet{Read2017}, and unlike \citealp{Behroozi2019} and \citealp{Stefanon2021}. Adding baryons to the virial masses would bring our simulations closer to the observational $z\simeq 5$ SMHM relations, by increasing $M_{\rm vir}$ by a factor of $\sim 1.2$.}. Even if the SMHM relations from our simulations have roughly the same slope as observational estimates, this may show that our feedback models remain too weak to sufficiently regulate star formation at high redshift. However, this is mitigated by uncertainties when inferring stellar masses from observations, especially for galaxies at high redshifts. Compared to the local observations of low-mass galaxies from \citet{Read2017}, the SMHM relation from our simulations are in broad agreement for intermediate mass halos, and above the observational estimates for halo masses around $10^{10}\rm\, M_\odot$. Given the difficulty in assessing stellar masses at high-redshift, we consider the UV luminosity function to provide a more robust comparison of our simulations to observations.

\begin{figure*}
    \includegraphics[width=0.9\textwidth] {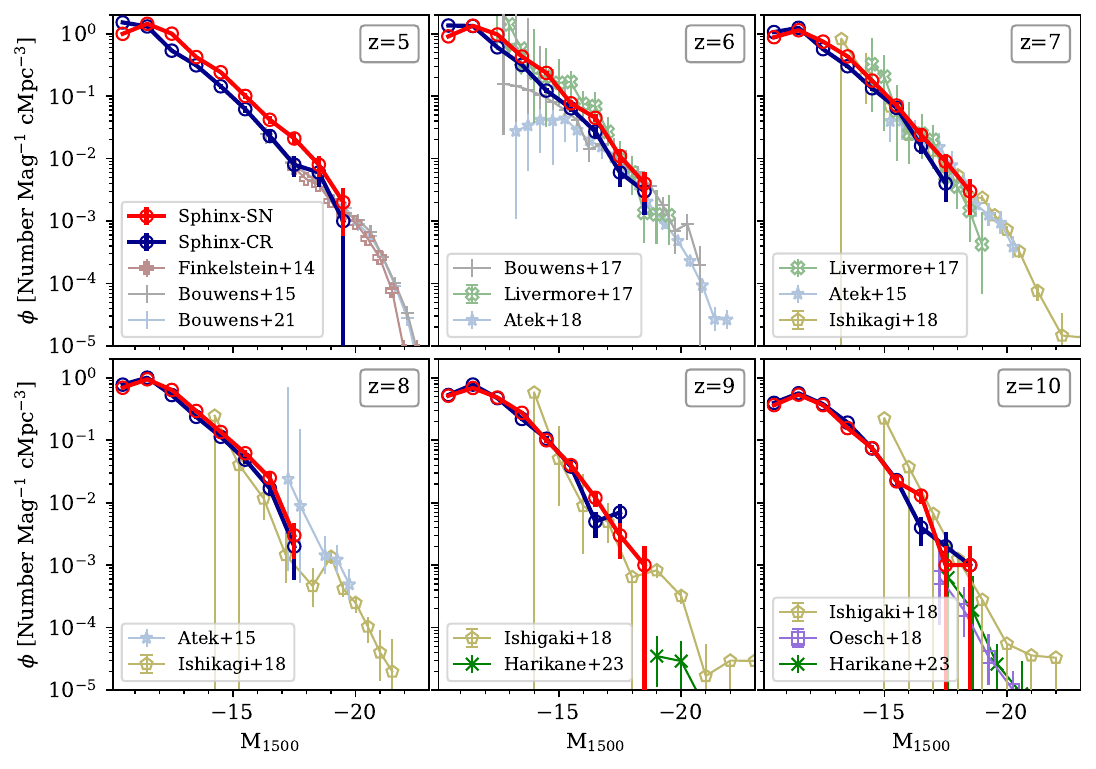}
    \centering
    \caption{Dust-attenuated UV luminosity function in \sphinxsn{} (red) and \sphinxcr{} (dark blue), with Poissonian error-bars. From the top left to the bottom right panel, we show increasing redshift from 5 to 10. The references of the observations shown in each panel are written in the legend. At any time, the UV luminosity functions of the two simulations are very similar and consistent with the observational constraints.}
    \label{fig:uvlum}
\end{figure*}

Figure~\ref{fig:uvlum} shows the dust-attenuated UV luminosity functions (UVLFs) for our two simulations, at redshift between 5 and 10 from the upper left to the bottom right panels. This provides another way to estimate the efficiency of feedback in regulating star formation. Unlike stellar mass that has to be inferred from SED model fitting, UV luminosity is a direct observable, which traces light emission from young and massive stars. To derive the UV luminosity of the simulated galaxies, we follow the procedure explained in Sect.~\ref{subsection:muv}. The UVLF thereby depicts the number of galaxies per volume in a given bin of $M_{1500}$. The observed UVLFs shown at different redshifts are taken from \citet{Finkelstein2015,Bouwens2015,Bouwens2017,Bouwens2021,Livermore2017,Atek2018,Ishigaki2018,Oesch2018,Harikane2023}. Observations are limited by the sensitivity of the instruments, and cannot accurately measure the faint end of the UVLF apart from a few cases where lensing magnification is utilised, in a small set of lensed fields. Conversely, the volume of our \sphinx{} simulations is too small to contain massive and bright galaxies, and therefore is limited to magnitudes fainter than -20. 

Overall, the two \sphinx{} simulations with and without CRs have similar UVLFs at any time. With decreasing redshift, \sphinxcr{} tends to have a lower UVLF at any magnitude. This is despite the \crs{} model being less efficient at suppressing star formation than the \nocr{} feedback in the most massive galaxies (Fig.~\ref{fig:smhm}). As the intrinsic UVLFs are barely distinguishable between the two simulations (Fig.~\ref{fig:uvlum_intr}), the slow drop of the UVLF towards low redshift with CRs is therefore a result of dust absorption. While the \nocr{} model is efficient at ejecting gas and metals out of massive galaxies, the \crs{} feedback acts differently by pushing gas more gently, which also traps metals, and thus dust, in the ISM. In \sphinxcr{}, massive galaxies have a gas and metal rich core (Fig.~\ref{fig:maps}), which leads to larger opacities and dust absorption than in \sphinxsn{}. In any case, the agreement between observations and the two \sphinx{} simulations indicates a realistic emission of UV radiation in our simulated galaxies. This result therefore shows that our model of CR feedback produces galaxies whose stellar mass and UV luminosity reasonably matches observational estimates.

\subsection{Global impact of CR feedback on reionisation}
\label{subsection:eor}

Now that we have established that our two \sphinx{} simulations similarly regulate star formation, we will show that they also lead to a similar intrinsic production of LyC photons, before showing how CRs impact the LyC escape fractions and the reionisation of the Universe. 

\begin{figure}
        \includegraphics[width=\columnwidth]{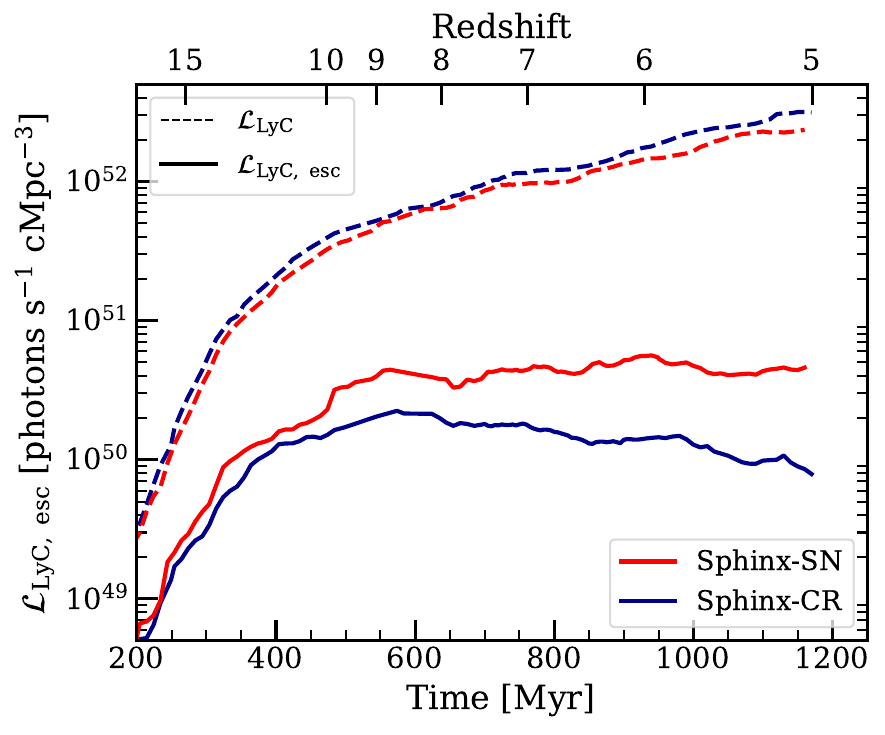}
        \centering
    \caption{Time evolution of the intrinsic LyC luminosity per volume $\mathcal{L}_{\rm LyC,}$ (dashed lines) and of the escaping LyC luminosity per volume $\mathcal{L}_{\rm LyC,esc}$ (solid lines). $\mathcal{L}_{\rm LyC,}$ and $\mathcal{L}_{\rm LyC,esc}$ are luminosity-weighted mean properties over the last 100 Myr, to smooth their bursty fluctuations with time. Even if the intrinsic LyC luminosities are similar in the two simulations, the escaping LyC luminosity is lower in \sphinxcr{}.}
    \label{fig:lumesc}
\end{figure}

In Fig.~\ref{fig:lumesc}, we first show the time evolution of the LyC luminosity emitted per unit volume $\mathcal{L}_{\rm LyC}$ (in dashed lines) and the escaping LyC luminosity per unit volume $\mathcal{L}_{\rm LyC,esc}$ (in solid lines), defined as the product of $\mathcal{L}_{\rm LyC}$ and the escape fraction $f_{\rm esc}$. Both quantities are averaged over 100 Myr. This demonstrates that while the number of intrinsically emitted LyC photons is roughly the same in the two simulations, the amount of LyC photons ionising the IGM is much lower with CR feedback (by up to a factor 6), due to reduced escape fractions.

\begin{figure}
        \includegraphics[width=\columnwidth]{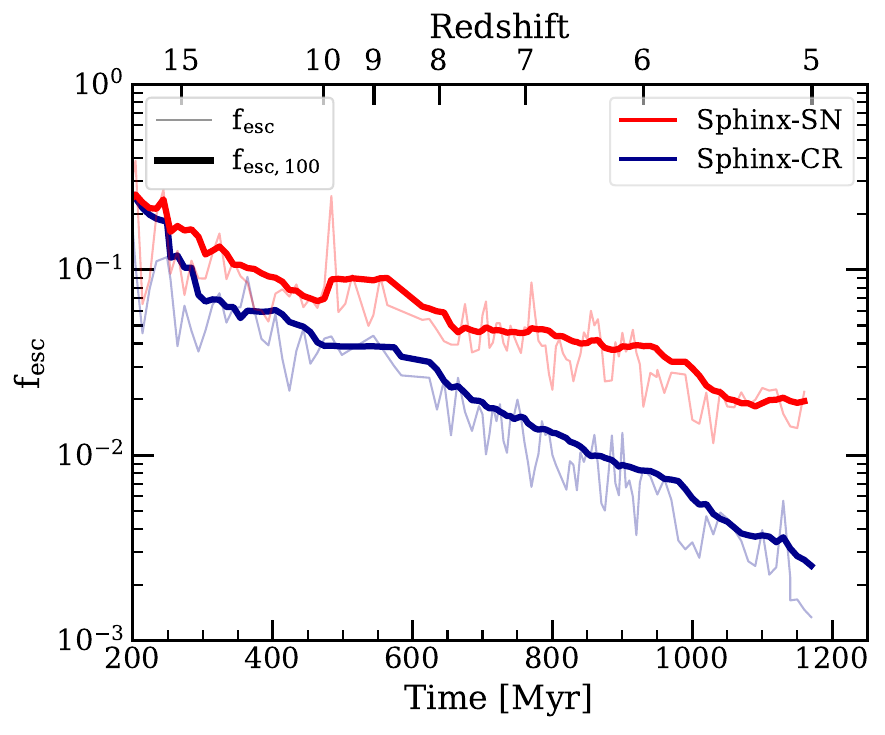}
        \centering
    \caption{Evolution with time of the global luminosity-weighted escape fraction of LyC photons $f_{\rm esc}$ in \sphinxcr{} (dark blue) and \sphinxsn{} (red). The luminosity-weighted mean escape fraction $f_{\rm esc}$ is shown in transparent lines, and the thick lines show the average over the last 100 Myr, to smooth the bursty fluctuations with time. In \sphinxcr{}, the escape fraction of ionising radiation is lower than in \sphinxsn{}.}
    \label{fig:fesc}
\end{figure}

We illustrate this in Fig.~\ref{fig:fesc}, which shows how the global luminosity-weighted escape fraction of LyC photons evolves with time in the two simulations. The escape fractions of LyC photons are lower in \sphinxcr{} than in \sphinxsn{}. The values differ by up to a factor of 7.7 at $z=5$, with a difference which tends to decrease with increasing redshift. In addition, the decrease of escape fraction with time is steeper in \sphinxcr{}.

\begin{figure}
        \includegraphics[width=\columnwidth]{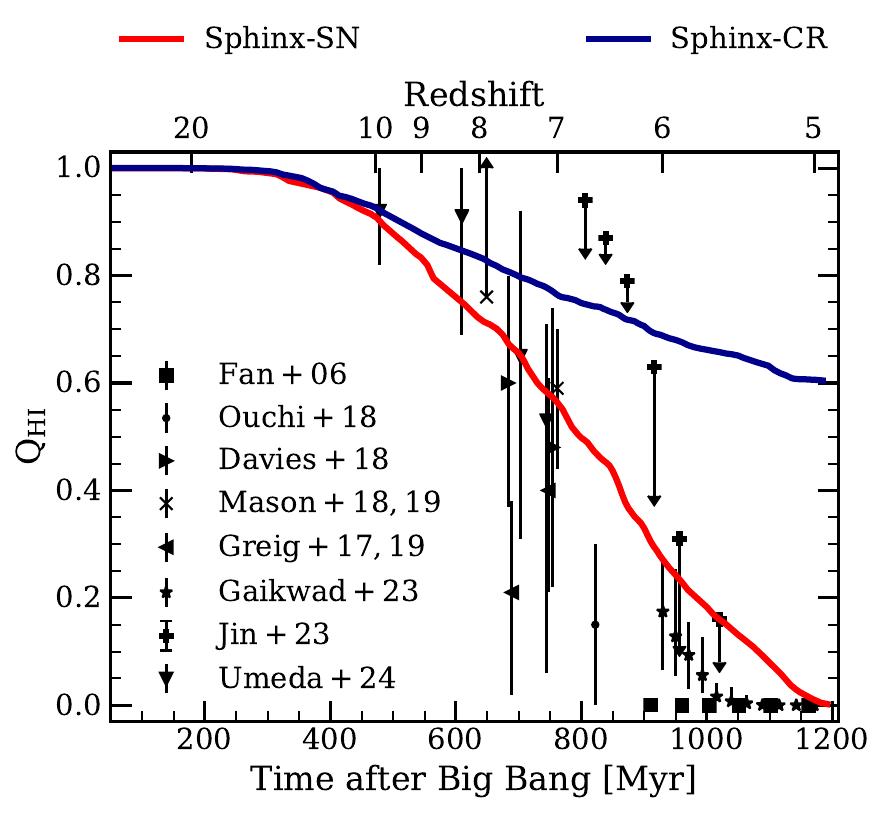}
        \centering
    \caption{Total volume-weighted fraction of neutral gas as a function of time in \sphinxsn{} (red) and \sphinxcr{} (dark blue). We additionally show with black data points observational estimates from studies indicated in the legend and in the text. The reionisation history is drastically delayed with CRs, and the simulation volume is still composed of 60 per cent of neutral hydrogen at $z=5$. The simulation volume of \sphinxsn{} is largely ionised by $z=5$, which is in much better agreement with the observational estimates.}
    \label{fig:QHI}
\end{figure}

We now consider how the different escape fractions affect the reionisation of the simulated volumes. Figure~\ref{fig:QHI} shows the volume-weighted fraction of neutral gas in the whole simulation volume as a function of time, comparing \sphinxsn{} and \sphinxcr{} to black data points corresponding to observational estimates from \citet{Fan2006,Ouchi2018,Davies2018,Mason2018,Mason2019a,Greig2017,Greig2019,Jin2023,Gaikwad2023, Umeda2024}. \sphinxsn{} is in relatively good agreement with the observational estimates, producing a realistic reionisation history. After $z=6$, the fraction of neutral hydrogen is very low ($Q_{\rm HII}\simeq10^{-2}$ at $z=5$) and the whole simulation volume can be considered ionised. However, the picture becomes completely different with CR feedback. After $z=10$, the fraction of neutral gas in \sphinxcr{} starts to diverge with that of \sphinxsn{} and remains much higher at any time. Between $z=7$ and $z=6$, \sphinxcr{} appears in better agreement with the data points from \citet{Jin2023}. However, we have to note that the latter are upper limits on the fraction of neutral hydrogen, determined from the fraction of dark pixels in the Lyman $\alpha$ and $\beta$ forests, which tends to favour a late reionisation scenario \citep{Zhu2021}. Otherwise, and especially between $z=6$ and $z=5$, the neutral gas fraction in \sphinxcr{} is well above the observational, and much more robust, estimates. At $z=5$, around 60 per cent of the \sphinxcr{} volume is still not ionised, which is in strong disagreement with observations. In Sect.~\ref{section:disc} we discuss the possible reasons for the escape fractions being too low in \sphinxcr{}.

\subsection{Effect of CR feedback on the escape of LyC photons as a function of halo mass and distance}
\label{subsection:fesc}

\begin{figure}
        \includegraphics[width=\columnwidth]{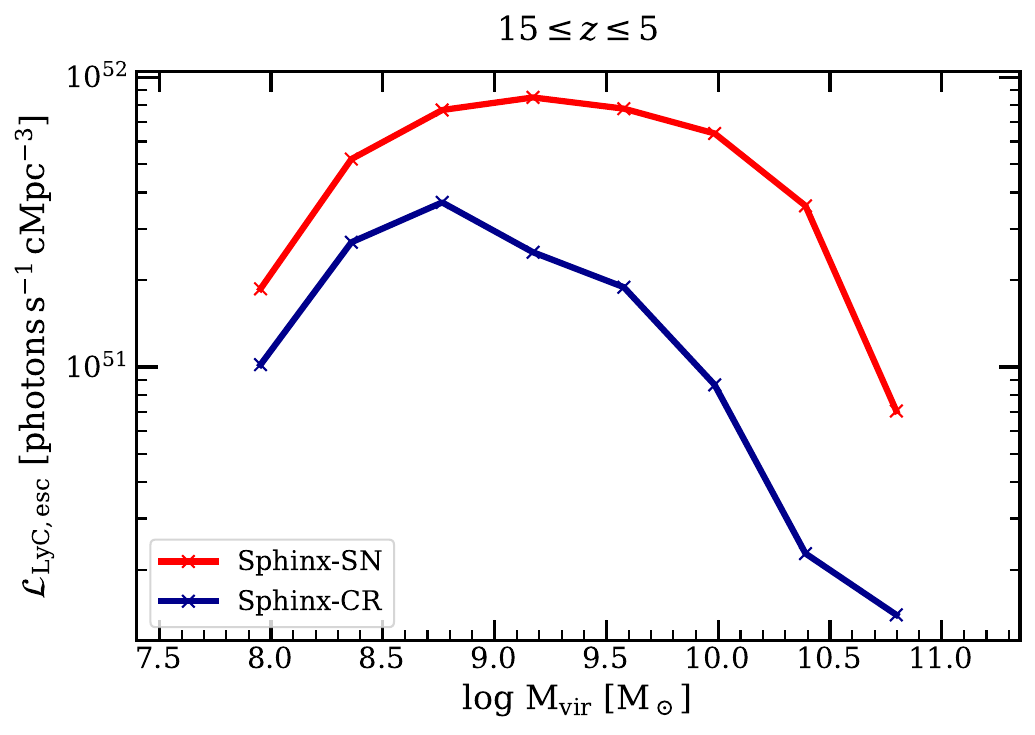}
        \centering
    \caption{Total escaping LyC luminosity per unit volume per bin of virial mass in \sphinxsn{} (red) and \sphinxcr{} (dark blue), for data stacked between $z=15$ and $z=5$. At any halo mass, the escaping LyC luminosity $\mathcal{L}_{\rm LyC,esc}$ is lower in \sphinxcr{}.}
    \label{fig:esclum_mvir}
\end{figure}

To determine which halos predominantly contribute to the reionisation process, Fig.~\ref{fig:esclum_mvir} shows the total escaping LyC luminosity per unit volume $\mathcal{L}_{\rm LyC,esc}$ between $z=15$ and $z=5$, as a function of halo mass. This allows us to assess the mass regimes that provide the bulk of escaping LyC photons. In \sphinxsn{}, halos with $8.5\leq\log(M_{\rm vir}/{\rm M_\odot})\leq9.8$ provide the highest total $\mathcal{L}_{\rm LyC,esc}$. In \sphinxcr{}, $\mathcal{L}_{\rm LyC,esc}$ is lower at any halo mass, and the emission of the bulk of escaping LyC photons is a bit more skewed towards lower mass halos. The difference in $\mathcal{L}_{\rm LyC,esc}$ between the two runs increases with increasing halo mass, with halos more massive than $10^{9}\rm\, M_\odot$ having escaping LyC luminosities up to 16 times lower with CR feedback than without. 

\begin{figure}
        \includegraphics[width=\columnwidth]{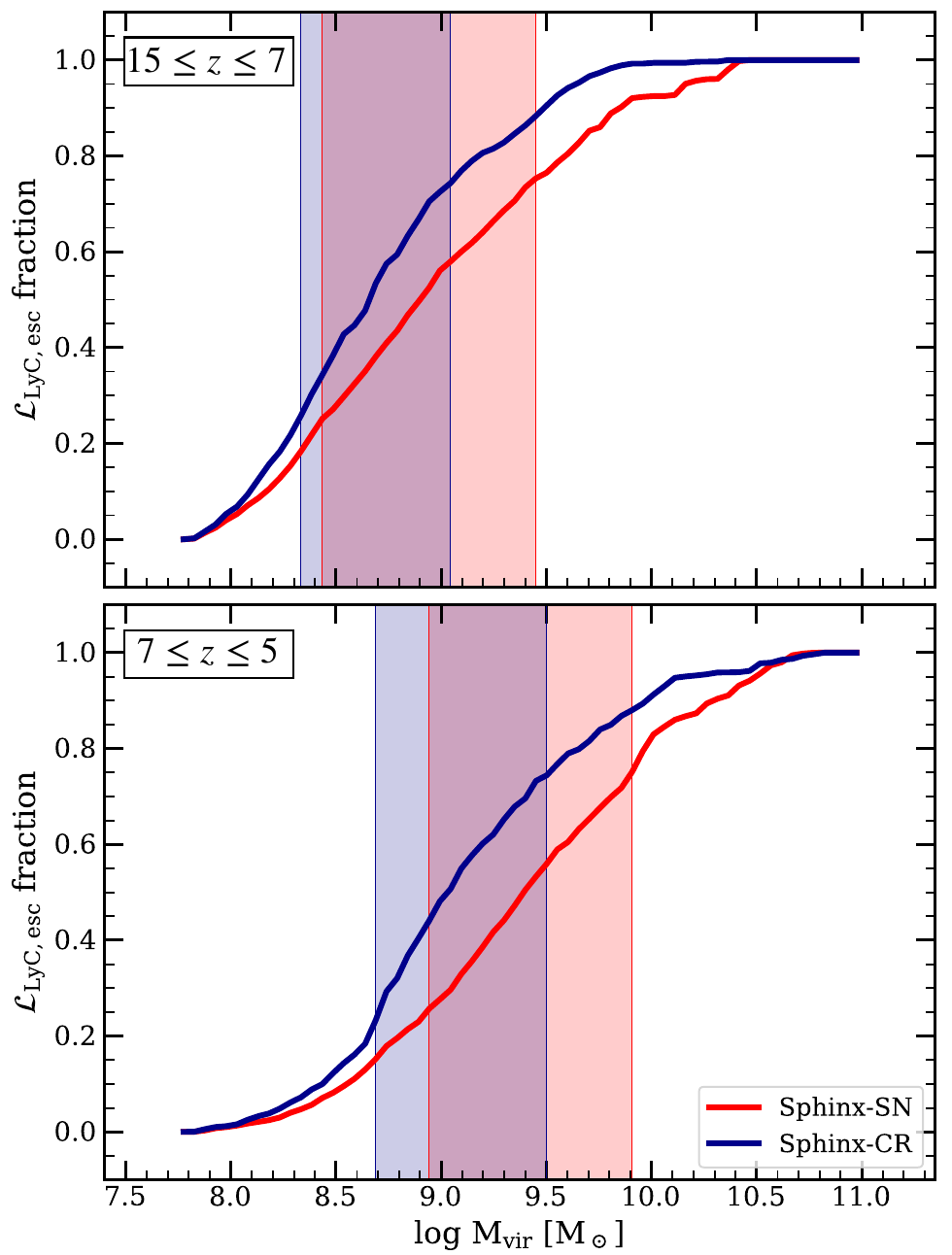}
        \centering
    \caption{Cumulative fraction of escaping LyC photons per bin of virial mass in \sphinxsn{} (red) and \sphinxcr{} (dark blue), for data stacked between $z=15$ and $z=7$ (upper panel) and between $z=7$ and $z=5$ (lower panel). The shaded areas show the lower and upper masses for which halos respectively contribute to 25\% and 75\% of the total escaping LyC photon budget at the corresponding redshift range. Lower mass halos contribute the most to reionisation, especially in \sphinxcr{}.}
    \label{fig:esclumcum_mvir}
\end{figure}

To determine more precisely the range of masses that contribute the most to the reionisation of the IGM, Fig.~\ref{fig:esclumcum_mvir} shows, as a function of virial mass, the cumulative escaping LyC luminosity between $z=15$ and $z=7$ (upper panel) and between $z=7$ and $z=5$ (lower panel), normalised by the total intrinsic LyC luminosity emitted during these redshift ranges. We enclose in shaded area the lower and upper mass limits for which halos contribute respectively to 25 and 75 per cent of the reionisation budget. At both redshift ranges, these limits are shifted towards lower masses in \sphinxcr{}, as a result of significantly lower $\mathcal{L}_{\rm LyC,esc}$ than in \sphinxsn{} for halos with $\log(M_{\rm vir}/{\rm M_\odot})\geq9$. Even if reionisation is not complete in \sphinxcr{}, we can conclude that the main drivers of the reionisation of the IGM in both simulations are low-mass halos with $8.5\leq\log(M_{\rm vir}/{\rm M_\odot})\leq9.8$ \citep[see also e.g.][]{Lewis2020,Kannan2022}. To give an idea of its representativeness, this mass range corresponds to more than 40 percent of all resolved halos at $z=5$ (see also Fig.~\ref{fig:hist}).

The difference of reionisation histories in our two simulations is the consequence of how feedback regulates the escape of ionising radiation. We have shown that CR feedback globally lowers the escape fraction of LyC photons. Now we quantify how this effect correlates with halo mass, and determine at which scales the LyC photons are primarily absorbed. 

\begin{figure}
        \includegraphics[width=\columnwidth]{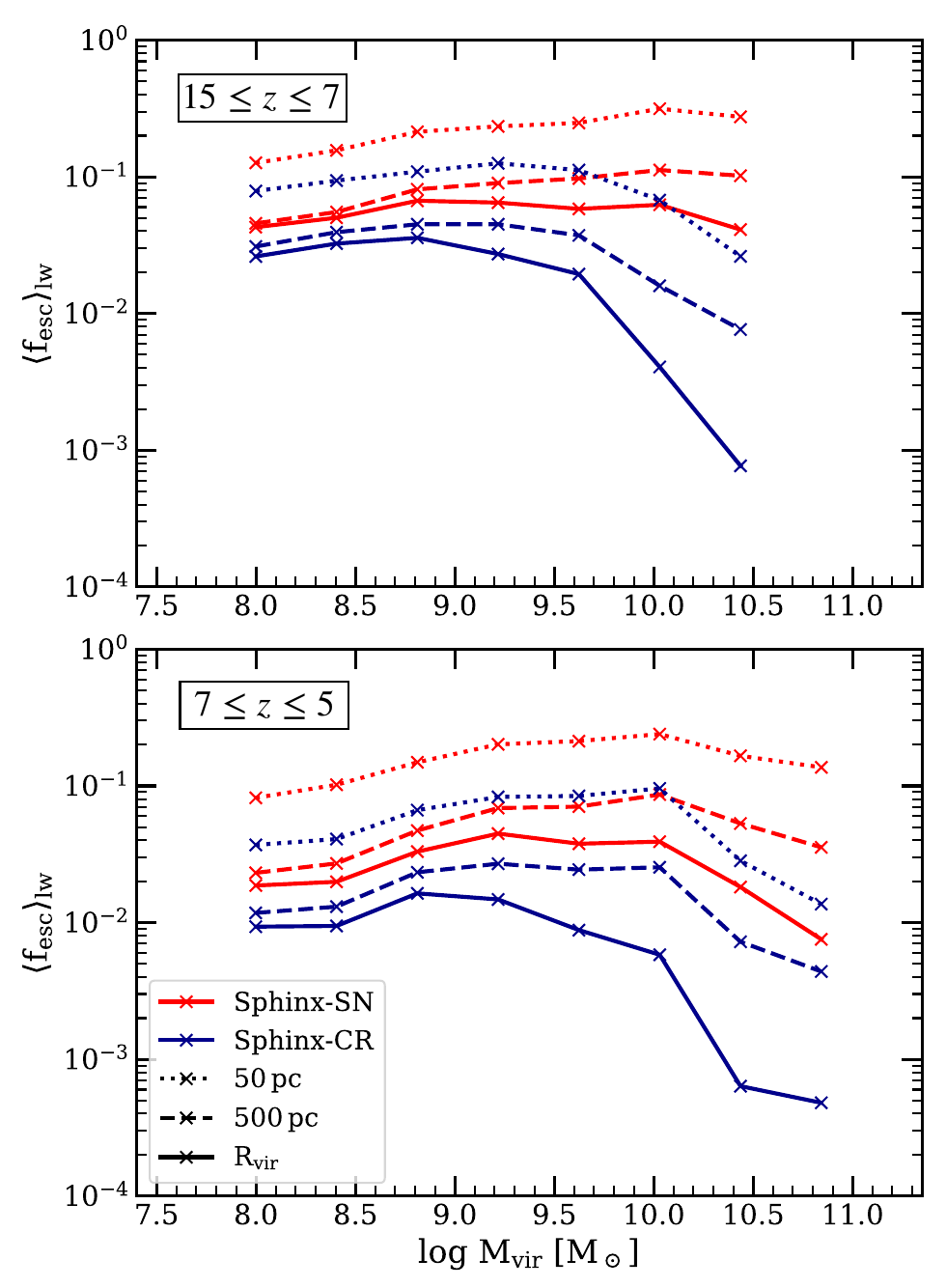}
        \centering
    \caption{Mean luminosity-weighted escape fraction per bin of virial mass in \sphinxsn{} (red) and \sphinxcr{} (dark blue), for data stacked between $z=15$ and $z=7$ (upper panel) and between $z=7$ and $z=5$ (lower panel). Dotted, dashed and solid lines respectively show $f_{\rm esc}$ at $50$ and $500\,\rm pc$ from the star particles and at the virial radius of the halos. On average, the escape fractions are smaller with CRs, especially for the most massive halos. In both simulations, most of the photons are absorbed in the close vicinity of the stars.}
    \label{fig:fesc_mvir}
\end{figure}

First we consider at which halo masses CR feedback mostly suppresses the escape of ionising radiation. Figure \ref{fig:fesc_mvir} shows the luminosity-weighted mean escape fraction $\langle f_{\rm esc}\rangle_{\rm lw}$ as a function of virial mass for data stacked between $z=15$ and $z=7$ (upper panel) and between $z=7$ and $z=5$ (lower panel). We first focus on the solid lines, that correspond to $f_{\rm esc}$ measured at the virial radii of the halos.

In \sphinxsn{}, $\langle f_{\rm esc}\rangle_{\rm lw}$ does not vary much with mass at $15\leq z\leq7$, and only decreases for the most massive halos at $7\leq z\leq5$ \citep[which is consistent with results from][]{Rosdahl2022}. In \sphinxcr{}, this decreasing trend with halo mass is more pronounced at both redshift ranges. We note that the very rightmost bin of $M_{\rm vir}$ may not be representative of the effect of CR feedback on 'massive' halos, as we are only focussing on a handful of objects (see Fig.~\ref{fig:hist}). Depending on the halo mass, the \crs{} feedback reduces the escape fraction of LyC photons compared to the \nocr{} feedback by a factor of between 1.5 and 54 at $15\leq z\leq7$, and of between 2 and 28 at $7\leq z\leq5$. Interestingly, we find that CRs reduce more significantly the fraction of escaping photons compared to the \nocr{} model in massive galaxies. This also explains the difference between the two simulations in Fig.~\ref{fig:esclum_mvir} and \ref{fig:esclumcum_mvir}, which is due to the escape fractions going down with halo mass in \sphinxcr{}. We will explain this behaviour later in the section.

We have extensively discussed the fact that the escape fraction of ionising photons is reduced in \sphinxcr{}. We now want to probe at which length scales CR feedback causes radiation to be increasingly absorbed. So far, the values of escape fractions shown were estimated at the virial radii of the halos. In order to determine if CR feedback prevents the escape of radiation in the ISM of the galaxies or further away, we compute the $f_{\rm esc}$ of LyC photons at arbitrary distances of $50\,\rm pc$ and $500\,\rm pc$ from the stellar particles, using the \rascas{} code and the same method as described in Sect.~\ref{subsection:rascas}. The distance of $50\,\rm pc$ is within the ISM of most galaxies (and is resolved at any time by at least 5 cell widths), and we chose $500\,\rm pc$ as an intermediate distance approaching the CGM (which is within the virial radius of all halos down to $z=15$). At $z=7$, the average gas density at $50\,\rm pc$ from the stellar particles is $28\,\rm H\,cm^{-3}$ in \sphinxsn{}, and it is 26 times higher in \sphinxcr{}. At $500\,\rm pc$ from the stars, this goes down to $1.5\,\rm H\,cm^{-3}$ and $14\,\rm H\,cm^{-3}$ in the runs without and with CRs, respectively. At the virial radius, the average gas densities are almost identical and equal to $4\times10^{-3}\,\rm H\,cm^{-3}$.

Each panel of Fig.~\ref{fig:fesc_mvir} shows $\langle f_{\rm esc}\rangle_{\rm lw}$ at $50\,\rm pc$ and $500\,\rm pc$ from the stellar particles in dotted and dashed lines. Naturally, the escape fractions decrease with distance, because the probability for a LyC photon to be absorbed increases with the amount of matter it goes through. The maximum escape fraction at $50\,\rm pc$ at both epochs in \sphinxsn{} is about 25 percent, and significantly lower in \sphinxcr{}. Regardless of CR feedback, this means that most of the LyC photons emitted are absorbed locally in the ISM of galaxies, close to the stellar particles that emit them. This has previously been found by for instance \citet{Kimm&Cen2014,Paardekooper2015,Trebitsch2017}. In addition, LyC photons escape less efficiently in \sphinxcr{} than in \sphinxsn{}. This may be attributed to the SN feedback in \sphinxsn{} being more efficient at disrupting star-forming clouds and their local environment, which makes it easier for photons to escape. More quantitatively, about two times more LyC photons escape this $50\,\rm pc$ limit in \sphinxsn{} than in \sphinxcr{}, and this goes up to a factor of ten for the most massive halos. 

This increased suppression of $f_{\rm esc}$ with halo mass can be explained as follows. In \sphinxsn{}, SN feedback is efficient enough to disrupt the ISM and provides the necessary condition for radiation to escape from galaxies. In \sphinxcr{}, SN feedback is weaker, and may be insufficient to act in a similar way in massive galaxies. In addition, CR pressure gently fills the channels carved by SN explosions, which increases the ISM hydrogen column densities and suppress $f_{\rm esc}$ at any halo mass. This can be linked back to the average gas density measured at $50\,\rm pc$ from the stars, which is consistently higher in \sphinxcr{}, increasingly absorbing the ionising radiation emitted. Assuming a constant CR diffusion (as we do), CRs need more time to diffuse out of the ISM of large, massive galaxies than of low-mass ones. During this time, the CR pressure gradient builds up and reduces the benefit of SN explosions in clearing the way for radiation to escape, which increasingly impedes the escape of LyC photons.

More generally, the difference in escape fractions between \sphinxsn{} and \sphinxcr{} increases with both halo mass and time, independently of where $\langle f_{\rm esc}\rangle_{\rm lw}$ is measured. Between $50\,\rm pc$ and $500\,\rm pc$ from the stellar particles, the escape fractions are similarly reduced in the two simulations with, on average, an optical depth 1.5 times higher at $500\,\rm pc$ than at $50\,\rm pc$ for both simulations. Therefore, the difference in escape fractions between the two simulations is mostly determined by the absorption of LyC photons in the first $50\,\rm pc$. In addition, in halos less massive than $10^{9}\rm\, M_\odot$, the fraction of photons propagating between $500\,\rm pc$ and $R_{\rm vir}$ is almost the same at both redshift ranges and in both simulations. However, the suppression of $f_{\rm esc}$ beyond $500\,\rm pc$ increases with halo mass, showing that LyC photons are further absorbed in the CGM of the most massive galaxies. For halos more massive than $10^{10}\rm\, M_\odot$, the average increase of optical depth between $500\,\rm pc$ and $R_{\rm vir}$ is a factor of 1.42 in \sphinxcr{}, and 1.35 in \sphinxsn{}. The suppression of $f_{\rm esc}$ in the CGM of massive galaxies is hence stronger in \sphinxcr{}, which likely is the consequence of the presence of dense CR-driven winds. On average, less than one photon every 1000 emitted manages to reach the virial radius boundary of massive halos in \sphinxcr{}, making their contribution (in this simulation) to the reionisation budget negligible.

\section{Discussion}
\label{section:disc}

By including CR feedback in \sphinxcr{}, we have sought to improve the feedback modelling of the \sphinx{} simulations, by increasing their physical accuracy. However, our \crs{} feedback model leads to incomplete reionisation. In this section we discuss the possible reasons why CR feedback prevents reionisation from taking place in \sphinxcr{} (Sect.~\ref{subsec:howto}). We then summarise some of the numerical and physical limitations of our work (Sect.~\ref{subsec:limits}).

\subsection{How to reconcile CR feedback and reionisation}
\label{subsec:howto}

Our results suggest a tension between CR feedback and the fact that the Universe reionised around $z\simeq 6$. Naively, this may come from either too low LyC luminosity, or too inefficient escape of LyC photons. Regarding the first argument, we checked (Fig.~\ref{fig:xi_ion}) that the redshift evolution of the ionising photon production efficiencies in our \sphinx{} simulations is compatible with measurements from recent observations \citep{Simmonds2024b}. However, we may fail in capturing the dependency of the ionising photon production efficiency with galaxy properties. The ionising photon rate production depends on the shape of the galaxy ionising spectrum, and therefore fluctuates with dust extinction and stellar properties (such as their age, mass, metallicity, the fraction of binary stars) which all vary with redshift \citep{Matthee2017,Matthee2022,Shivaei2018,Chisholm2019,Atek2022}. In addition, this study focuses exclusively on ionising radiation produced by stars, and neglects LyC nebular emission \citep{Simmonds2024a} and X-ray binaries \citep{Mirabel2011,Eide2020}, as well as any contribution from quasars and active galactic nuclei (AGNs). We note, however, that the latter likely plays a role in galaxies more massive than those produced in our $\left(10\,\rm Mpc\right)^3$ simulations \citep{Hopkins2006,Reines&Volonteri2015,Zeltyn&Trakhtenbrot2022} and are thought to be too rare to dominate the UV background before $z=4$ \citep{Onoue2017,Trebitsch2021}. It is also worth noting that we made no adjustment to the SED, while using different SN feedback strengths in our two simulations. This choice was made in order to better isolate and focus on the effect of CR feedback. However, it is known that different SEDs can also impact the UV luminosity functions and, via the LyC photon budget, reionisation \citep{Ma2016,Rosdahl2018,Gotberg2020}.

The UV luminosity functions at high redshift in \sphinxcr{} are in fair agreement with observations, and we have shown that the different reionisation histories between our two simulations is due to the lower escape fractions with CRs. Although we find that CR feedback is too suppressive for the escape of ionising radiation, CRs are present in the real Universe, and the hydrogen in the IGM did reionise successfully around a billion years after the Big Bang. Therefore, one possible explanation for having too low escape fractions in \sphinxcr{} is that our calibrated \crs{} feedback model is not representative of the behaviour of the CR population in the young Universe. Only a few observations in our very local Universe exist to constrain the amount and escape time of CR energy within star-forming galaxies. One of the main and only ways to estimate the accuracy of the modelling of CRs is to rely on $\gamma$-ray luminosities, which correlate with the SFRs of galaxies. A higher SFR implies more SN explosions, and so more CR energy injection. Depending on CR diffusivity, this non thermal energy interacts with the ISM gas, emitting $\gamma$-ray photons as a result of hadronic collisions \citep{Guo&Oh2008}. If too much CR energy is injected or if it interacts for too long with the ISM, the reconstructed $\gamma$-ray emission will be too high compared to what is observed. We have checked (but do not show) that the $\gamma$-ray luminosities in \sphinxcr{} are an order of magnitude higher than estimates from local starburst galaxies \citep{Chan2019,Arturo2022,Fermi2022}. Based on these present-day observations, this may indicate that we are injecting too much CR energy, or that it interacts with the ISM for too long (i.e. with a too slow CR diffusion). As a result from our calibration strategy, we may overestimate the contribution of CRs in regulating star formation and impacting the galactic gas distribution, which globally slows down reionisation. This also means that CR feedback cannot be regarded as the only physical substitute to the boosted SN feedback used in \sphinx{}, and that other important physical processes, or better resolution, are needed to capture the regulation of galaxy growth (see Sect.~\ref{subsec:limits}). 

\subsection{Numerical and physical limitations}
\label{subsec:limits}

CR feedback is predominantly determined by the efficiency of CR energy injection and by its transport, parametrised with a diffusion coefficient. In this study, the values of these two parameters (i.e. the fraction of SN energy released in CRs and the diffusion coefficient) are constant and, even if chosen to be in the realm of acceptable values, likely to be an oversimplification of reality. Recently, efforts have been made to account for a better modelling of CR transport in simulations of galaxies, with a diffusion coefficient varying with local plasma properties \citep{Farber2018,Semenov2021,Hopkins2021c}. A spectral implementation of CR transport model has also been developed by \citet{Girichidis2022}, which resolves the CR energy spectrum from $\sim$ MeV to TeV, and models the diffusion coefficient and energy losses depending on the energy of the CR population (see also~\citealp{Hopkins2022}). These prescriptions intrinsically impact the coupling and the dynamical impact of CRs at ISM scales, likely affecting the escape of ionising radiation. With all of the more realistic approaches aforementioned (some of them being computationally expensive), it would be particularly interesting to quantify how CR feedback differently regulates the reionisation process. As a starting point, \citet{Farcy2022} showed in idealised galaxy simulations that faster CR diffusion reduces the impact of CR feedback in suppressing the escape of ionising photons. In our planned follow-up paper, we will explore the impact of CR energy injection and diffusion on star formation and reionisation, using a set of smaller \sphinx{} volumes.

Another aspect of CR transport that is neglected in this study is CR streaming, that originates from the interaction of CRs with Alfvén waves \citep{Kulsrud&Pearce1969}. By modelling the scattering of CRs off of Alfvén waves and the associated damping processes, \citet{Thomas2024} report a decoupling of CRs at ISM scales in their idealised galaxy simulations. This leads to a similar porosity of the ISM in their simulations with and without CRs, which would reduce the impact of CRs on the small-scale escape of LyC photons. Using zoom simulations of a dwarf galaxy \citep{Martin-Alvarez2023}, \citeauthor{Yuan2024} (\citeyear{Yuan2024}, and also \citealp{Yuan2024b}, using a more massive galaxy) showed that including CRs and CR streaming produces the strongest feedback compared to SN and radiation alone \citep[but see][]{Dashyan&Dubois2020,Chan2019}, allowing higher values of LyC escape fractions to be reached, in contrast to our findings. Compared to \sphinxcr{}, and in addition to CR streaming, their simulations use slightly different star formation and SN feedback models. They adopt the magneto-thermo-turbulent star formation model described by \citet{Martin-Alvarez2020}, which may impact the burstiness of star formation compared to the thermo-turbulent prescription used in this study \citep{Kimm2017,Trebitsch2017}. Per SN explosion, they inject a total energy of $2\times10^{51}\,\rm erg$, with 10 per cent in the form of CR energy, unlike $10^{51}\,\rm erg$ and 20 per cent in our case. We also have twice more SN explosions per $100\,\rm M_\odot$ formed in our simulation with CRs than in their zoom simulations. In work in progress, we are currently investigating which of these differences impacts the effect of CR feedback on escape fractions. This will contribute to reconcile CRs with reionisation, in order to better quantify the impact of CR feedback on galaxies and during the EoR.

In addition to the uncertainties and simplifications regarding CR transport, the details of SN feedback can also impact our conclusions. Both the numerical modelling of SN feedback and the clustering and timing of SN explosions affect the strength of SN feedback in generating galaxy-scale outflows \citep{Rosdahl2017,Keller&Kruijssen2022,Farcy2022}, and its ability to clear sight lines for radiation to escape \citep{Trebitsch2017}. In their \spice{} simulations, \citet{Bhagwat2024} extensively study the impact of the SN explosion times and energies, and interestingly show that their more physically motivated prescription for SN feedback does not produce a realistic reionisation history. Although their results support the importance of stellar feedback in modulating the escape fraction of ionising radiation, they also hint towards the complementary role of other physical processes and resolution. 

In this regard, \citet{Kimm2017} show that radiation feedback, rather than SN explosions, determine the escape of radiation in mini-haloes by disrupting star-forming clouds. However, to capture this process, they use a very high resolution of $0.7\,\rm pc$ that is beyond reach in non-zoomed cosmological simulations. This connects back to the importance of resolving the multiphase structure of the ISM, which are the scales at which the propagation of radiation is primarily impacted. Indeed, properties of the ISM, such as its porosity and clumpiness, have been shown to be important for regulating $f_{\rm esc}$ \citep{Clarke&Oey2002,Fernandez&Shull2011,Ma2015}, already at the scales of molecular clouds and HII regions \citep{Dale2012,Geen2015,Kimm2019,Kakiichi&Gronke2021,Kimm2022}.

Overall, the stochasticity and burstiness of star formation, and the subsequent time and space distribution of SN explosions, all have a dominant, but highly non linear effects \citep[e.g.][]{Keller&Kruijssen2022}, on both galaxy growth and reionisation. This makes it even harder to interpret the role of CR feedback and its interplay with the other physical processes involved in galaxy evolution. Even if our simulations include state-of-the art physical models of star formation, SN, radiation and CR feedback for cosmological simulations, they still lack a number of complementary physical mechanisms. We already mentioned that we do not include the contribution of AGN to the production of LyC photons. Given the large number of (low-luminosity) AGN observed at high redshifts \citep[e.g.][]{Matthee2024,Akins2024}, AGN feedback may also play a role in the escape of the hydrogen ionising radiation, even at the low-mass regime captured in our simulations (\citealp{Dashyan2018,Koudmani2022}, but see \citealp{Dubois2015} or \citealp{Trebitsch2018}). In particular, because CRs are accelerated at shocks, they can also be indirectly injected by AGN and have an important contribution on galaxy evolution \citep{Wellons2023}. While SN, AGN and CR feedback differently regulate galaxy evolution and the escape of ionising photons, their complex and non linear interplay may produce a different effect than when they are individually considered \citep{Biernacki&Teyssier2018}, and lead to different conclusions regarding the reionisation.

Finally, we suggest a few other improvements that would enhance the physical fidelity of our work. Because it is intimately related to stellar feedback, star formation and its subgrid modelling is another important aspect of EoR studies. In the first place, the process of star formation can be better translated with more sophisticated chemical networks that model molecular gas cooling \citep[e.g.][]{Kim2023}. In addition, in this work stars are modelled as stellar populations. Although it would be prohibitively expensive to resolve individual stars, capturing individual supernova explosions would improve our understanding of their effect on the escape fraction. As an intermediate approach, \citet{Kang2024} showed that modelling stars with sink particles leads to a bursty star formation and effective gas clump disruption, potentially also impacting the escape fractions.

Another promising avenue for improving both our inference of galaxy properties (such as the UV luminosity function), and of the propagation of LyC radiation, is to incorporate dust evolution in galaxy simulations. Here, we only apply a simplified model of dust absorption in post-processing \citep{Garel2021}, but recent developments have enabled the self-consistent modelling of dust growth, destruction, and evolution \citep{Dubois2024}. Using simplifying assumptions about the dust grain distribution, some cosmological simulations of the EoR already include dust models \citep{Trebitsch2021,Kannan2022,Lewis2023}, but the effect of dust on galaxy properties and on reionisation remains to be determined.

On a different note, we should also remind that the size of our \sphinx{} volume is too small for modelling galaxies in halos more massive than $10^{11}\,\rm M_\odot$. To date, the largest of the \sphinx{} simulations has a width of $20\,\rm cMpc$ (without CRs, \citealp{Rosdahl2022,Katz2023}). This intrinsically limits the size of the ionised regions in the IGM. It also prevents a deeper exploration of the impact of stellar feedback and its interplay with the escape of radiation in the brightest galaxies of the high-redshift Universe \citep{Naidu2022,Mason2023}. However, given the computational cost of the physics included in our simulations, we argue that the volume and resolution adopted currently remain our best compromise.

\section{Conclusions}
\label{section:ccl}

The study of the high-redshift Universe, from the formation and the evolution of the first galaxies to the reionisation of the inter-galactic medium, remains one of the main current scientific challenges. This study investigates the impact of CRs on the growth of high-redshift galaxies during the EoR and their contribution to the reionisation process. For this purpose, we performed two \sphinx{} cosmological RMHD simulations with and without CRs (respectively labelled \sphinxcr{} and \sphinxsn{}), using the \ramsesrt{} code \citep{Teyssier2002,Rosdahl2013,Rosdahl&Teyssier2015}. This has allowed us to combine, for the first time, non-equilibrium chemistry in a multiphase inter-stellar medium, radiative transfer of hydrogen and helium ionising radiation, mechanical SN feedback, and CR injection and transport in a $\left(10\,\rm cMpc\right)^3$ cosmological simulation, resolving more than 3000 star-forming halos down to $\sim 10\,\rm pc$ at $z=6$. In order to produce realistic stellar-to-halo mass relation and UV luminosity functions at high redshift, we used a \crs{} feedback model, by injecting 20 per cent of the SN energy into CRs and adopting a SN rate of $\rm 2\, SNe$ per $100\,\rm M_\odot$ of newly formed stars. This calibration reduces the tension with expectations for the rate of SN explosions from stellar population models ($\rm 1\, SN/100\, M_\odot$ for a \citealp{Kroupa2001}- or a \citealp{Chabrier2003}-like IMF) compared to the $4$-fold boosted SN rate used in the fiducial \sphinx{} simulations \citep{Rosdahl2018,Rosdahl2022}, that we refer to as the \nocr{} model. With this parametrisation, we find that our \crs{} feedback suppresses star formation in a similar way as the \nocr{} feedback, with a decreasing efficiency for the most massive halos that only slightly impacts the total stellar mass formed. More importantly, both simulations reproduce observations of the UV luminosity function fairly well at $z=5-10$. Our conclusions regarding the role of CRs on reionisation are given below: 

\begin{itemize}
\item The impact of CRs on reionisation is predominantly determined by their effect on the escape of LyC photons. Due to CRs and weaker SN feedback, the escape fraction of LyC photons is lower in \sphinxcr{} than in \sphinxsn{} at any halo mass. This reduction in $f_{\rm esc}$ is dominant in the close vicinity of the stars, with most of the ionising photons being absorbed within $50\,\rm pc$ of the sources that emit them. Compared to \sphinxsn{}, the fraction of photons that escape this $50\,\rm pc$ limit in \sphinxcr{} is decreased by a factor varying between 1.6 and 10.5, depending on the redshift and halo mass range considered. At any time and distance from the emitting sources, the escape fraction of LyC photons is  lowest for massive halos when CRs are included, making their contribution to the total escaping LyC radiation budget negligible. 

\item The \crs{} feedback leads to a delayed and incomplete reionisation. While both simulations have a similar intrinsic production of LyC radiation, including CR feedback reduces the global luminosity-weighted escape fraction of LyC photons at any time, compared to the case with strong SN feedback alone. In \sphinxsn{}, the neutral gas fraction in the IGM roughly agrees with observational estimates and the IGM becomes entirely ionised by $z=5$. On the other hand, \sphinxcr{} retains a total volume-weighted neutral gas fraction of 60 per cent at $z=5$, which is inconsistent with observations of the Lyman alpha forest.
\end{itemize}

Our study indicates that when CR feedback sufficiently regulates galaxy growth at high redshift, it also hinders the reionisation process. However, precisely quantifying these effects is challenging due to several uncertainties, such as the precise amount of CR energy, its transport in high-redshift galaxies, the impact of other complementary feedback mechanisms (none of which are currently included), and the limitations of sub-grid models to translate the complexity of physical processes acting at unresolved scales. This suggests the importance of improving our galaxy formation models by modelling feedback from first principles to capture the complexity of the reionisation process. In combination with high-redshift observations from current and upcoming revolutionary facilities such as JWST, ALMA, SKA, and ELT, this paves the way forward in gaining a more accurate understanding of the processes governing galaxy evolution through cosmic time.

\begin{acknowledgements}
      We gratefully thank Maxime Rey and Aniket Bhagwat for insightful discussions, and Léo Michel-Dansac for helping with the \rascas{} code. This work has been granted access to the HPC resources of TGCC under the allocation 2022-A0120413449 made by GENCI. We also acknowledge support from the CBPsmn (PSMN, Pôle Scientifique de Modélisation Numérique) of the ENS de Lyon for the computing resources. The platform operates the SIDUS solution \citep{psmn} developed by Emmanuel Quemener. MF acknowledges funding from the Swiss National Science Foundation (SNSF) via a PRIMA Grant PR00P2 193577 “From cosmic dawn to high noon: the role of black holes for young galaxies”. S.M.A. is supported by a Kavli Institute for Particle Astrophysics and Cosmology (KIPAC) Fellowship, and by the NASA/DLR Stratospheric Observatory for Infrared Astronomy (SOFIA) under the 08\_0012 Program. SOFIA is jointly operated by the Universities Space Research Association, Inc. (USRA), under NASA contract NNA17BF53C, and the Deutsches SOFIA Institut (DSI) under DLR contract 50OK0901 to the University of Stuttgart. TK is supported by the National Research Foundation of Korea (NRF) grant funded by the Korea government (2022R1A6A1A03053472 and RS-2022-NR070872) and by the Yonsei Fellowship funded by Lee Youn Jae.\\

\textsl{Software}: \texttt{Numpy} \citep{VanderWalt2011}, \texttt{Matplotlib} \citep{Hunter2007}, \ramses{} \citep{Teyssier2002}, \rascas{} \citep{rascas2020}

\end{acknowledgements}

\bibliographystyle{aa}
\bibliography{biblio}

\begin{appendix}
\onecolumn
\section{Intrinsic UV luminosity function}
\label{app:uvlf}

\begin{figure*}[h!]
    \includegraphics[width=0.75\textwidth]{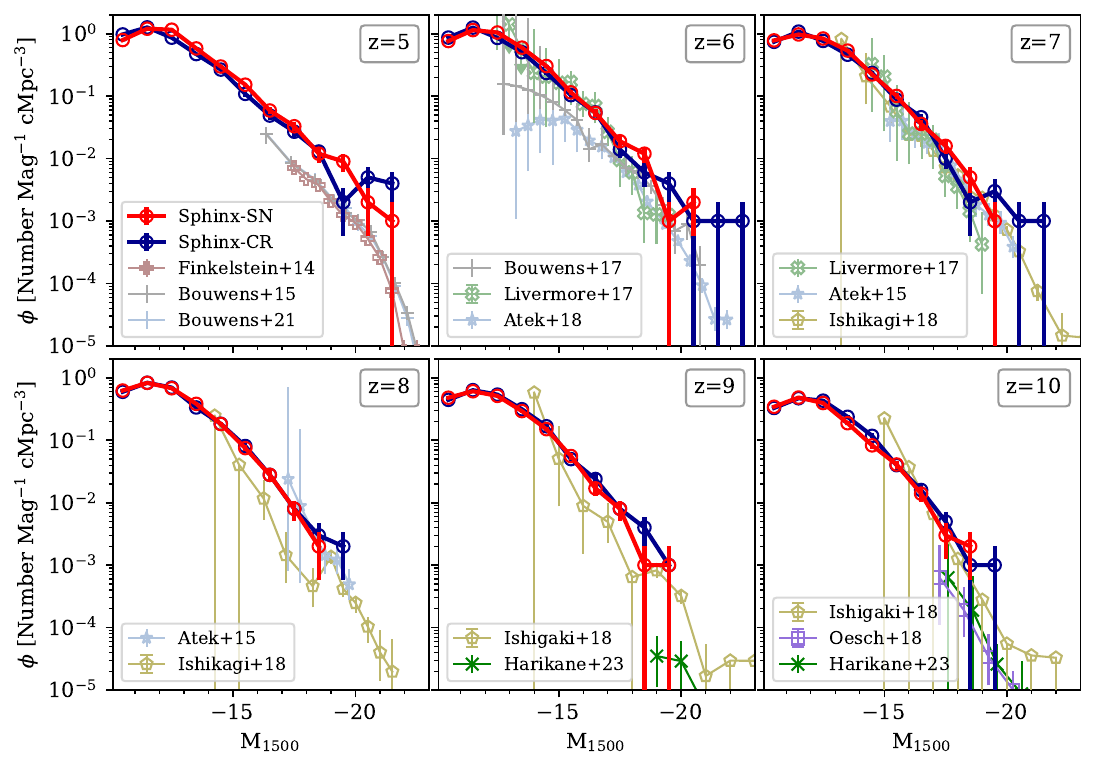}
    \centering
    \caption{Intrinsic UV luminosity function in the \sphinx{} simulations with and without CRs (in dark blue and red), with Poissonian error-bars. From the upper left to the lower right panel, we show increasing redshift from 5 to 10. The references of the observations shown in each panel are written in the legend.}
    \label{fig:uvlum_intr}
\end{figure*}

Figure~\ref{fig:uvlum_intr} shows the intrinsic UV luminosity functions for our two simulations from redshift $z=5$ to $z=10$. The intrinsic UVLF is brighter in \sphinxcr{} than in \sphinxsn{}, especially at low-redshift. This is consistent with the fact that the most massive galaxies simulated have higher stellar masses with the \crs{} feedback model than with the \nocr{} one (Fig.~\ref{fig:smhm}), thereby being intrinsically more luminous. By comparing to Fig.~\ref{fig:uvlum}, this also demonstrates that dust absorption is particularly relevant in \sphinxcr{} for the brightest galaxies, and at decreasing redshift.

\section{Ionising photon production efficiency}
\label{app:xi_ion}

\begin{figure}[h!]
    \includegraphics[width=0.35\textwidth]{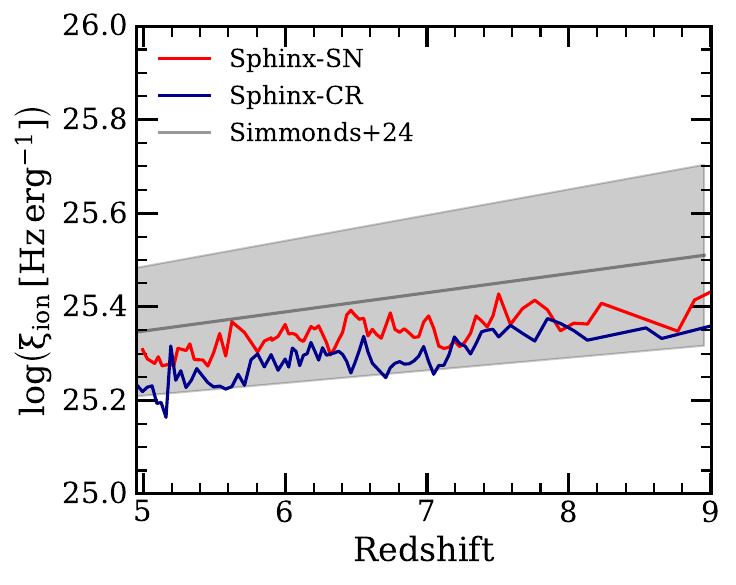}
    \centering
    \caption{Ionising photon production efficiency as a function of redshift in \sphinxcr{} (in dark blue) and \sphinxsn{} (in red). The grey line shows the best fit from spectroscopic data from \citet{Simmonds2024b}.}
    \label{fig:xi_ion}
\end{figure}

\begin{multicols}{2}

\noindent Figure~\ref{fig:xi_ion} shows the ionising photon production efficiency $\xi_{\rm ion}$ as a function of redshift, where $\xi_{\rm ion}$ is the ratio of the ionising photon production rate to the UV luminosity. The values are of the same order of magnitude for the two simulations, and tends to be lower in \sphinxcr{} because of its slightly higher intrinsic UV luminosity. In both cases, $\xi_{\rm ion}$ inferred from our simulations is in good agreement with the observational fit from \citet{Simmonds2024b}, using spectroscopic data at $3\leq z\leq9$ from the JWST advanced deep extragalactic survey.

\end{multicols}
\end{appendix}
\end{document}